\documentclass[AMA,STIX1COL]{WileyNJD-v2}

\usepackage{enumitem}
\usepackage[capitalise]{cleveref}
\usepackage{csquotes}
\usepackage{threeparttable}
\usepackage{colortbl}

\usepackage{booktabs}
\definecolor{Gray}{RGB}{100,100,100}
\definecolor{naturelight}{RGB}{248,231,215}
\definecolor{naturedark}{RGB}{204,139,136}
\definecolor{myblue}{RGB}{0,100,200}
\definecolor{myred}{RGB}{204,102,0}

\usepackage{siunitx}
\DeclareSIUnit\mmHg{mmHg}
\hypersetup{
	colorlinks,
	linkcolor=myred,
	citecolor=myblue,
	urlcolor=myblue
}

\articletype{Applied Research}%

\received{}
\revised{}
\accepted{}

\begin{document}

\title{An approach to study recruitment/derecruitment dynamics in a patient-specific computational model of an injured human lung}

\author[1]{Carolin M. Geitner*}

\author[2]{Tobias Becher}

\author[2]{In\'{e}z Frerichs}

\author[2]{Norbert Weiler}

\author[3]{Jason H. T. Bates}

\author[1]{Wolfgang A. Wall}

\authormark{GEITNER \textsc{et al}}

\address[1]{\orgdiv{Institute for Computational Mechanics}, \orgname{Technical University of Munich}, \orgaddress{\state{Garching b. Muenchen}, \country{Germany}}}

\address[2]{\orgdiv{Department of Anesthesiology and Intensive Care Medicine}, \orgname{University Medical Center Schleswig-Holstein, Campus Kiel}, \orgaddress{\state{Kiel}, \country{Germany}}}

\address[3]{\orgdiv{Department of Medicine}, \orgname{University of Vermont College of Medicine}, \orgaddress{\state{Burlington, Vermont}, \country{USA}}}

\corres{*Carolin M. Geitner, Institute for Computational Mechanics, Technical University of Munich, Boltzmannstrasse 15, 85748 Garching b. Muenchen, Germany. \\ \email{carolin.geitner@tum.de}}

\abstract[Summary]{%

We present a new approach for physics-based computational modeling of diseased human lungs. Our main object is the development of a model that takes the novel step of incorporating the dynamics of airway recruitment/derecruitment into an anatomically accurate, spatially resolved model of respiratory system mechanics, and the relation of these dynamics to airway dimensions and the biophysical properties of the lining fluid. The importance of our approach is that it potentially allows for more accurate predictions of where mechanical stress foci arise in the lungs, since it is at these locations that injury is thought to arise and propagate from. We match the model to data from a patient with acute respiratory distress syndrome (ARDS) to demonstrate the potential of the model for revealing the underlying derangements in ARDS in a patient-specific manner. To achieve this, the specific geometry of the lung and its heterogeneous pattern of injury are extracted from medical CT images. The mechanical behavior of the model is tailored to the patient's respiratory mechanics using measured ventilation data. In retrospective simulations of various clinically performed, pressure-driven ventilation profiles, the model adequately reproduces clinical quantities measured in the patient such as tidal volume and change in pleural pressure. The model also exhibits physiologically reasonable lung recruitment dynamics and has the spatial resolution to allow the study of local mechanical quantities such as alveolar strains. This modeling approach advances our ability to perform patient-specific studies in silico, opening the way to personalized therapies that will optimize patient outcomes.

}


\maketitle

\section{Introduction}
\label{sec:1}

Mechanical ventilation is a life-saving therapeutic measure for patients suffering from acute respiratory distress syndrome (ARDS), an often fatal condition that has recently gained widespread attention due to its association with severe cases of COVID-19~\cite{Li2020,Fan2020}. The main goal of mechanical ventilation is to maintain blood oxygenation and to ensure the removal of carbon dioxide from the lungs. At the same time, mechanical ventilation must be delivered in a manner that minimizes ventilator-induced lung injury (VILI) and its consequent exacerbation of organ damage~\cite{Gattinoni2003,Slutsky2013,Beitler2016}. Two prevailing mechanisms fostering VILI are (i) overdistension of the alveolar tissue, often referred to as \textit{volutrauma}, and (ii) cyclic opening and closure of unstable lung units, often referred to as \textit{atelectrauma}~\cite{Beitler2016,Bates2018}. 

Providing a patient with sufficient minute ventilation while simultaneously minimizing VILI is a challenging task for physicians. For the general ARDS case, various protective ventilation strategies have been developed to reduce the aforementioned injury risks~\cite{Slutsky2013} and to improve gas exchange in the lung. Such strategies include the reduction of tidal volume, the application of positive end-expiratory pressure (PEEP), and the use of recruitment maneuvers to open collapsed lung regions. Despite the established clinical benefits of these techniques~\cite{ARDS2000}, however, they do not take inter-patient variability into account. This is due in part to the fact that only global parameters of ventilation, gas exchange and respiratory mechanics are typically at hand. Such parameters do not provide insight into the distributed nature of lung injury in what is usually a very heterogeneous pathology. Being able to take both global and local measures of lung function into account when making treatment decisions would be of great benefit to the management of ARDS~\cite{Kollisch2018}. The clinical application of electrical impedance tomography (EIT) serves this purpose at the bedside to some extent by deducing regional ventilation distributions based on a map of the electrical properties of the thorax~\cite{Frerichs2017}. However, EIT is limited to a slice of the organ (2D-EIT) and has only very coarse resolution. Thus, at the present state of the art, (1)~a detailed and comprehensive insight (2)~during therapy (3)~into the whole lung is not available.

Computational models of the lung that represent ARDS pathophysiology in a personalized manner may thus be helpful in optimizing protective ventilation strategies in clinical practice, especially if they include descriptions of phenomena such as heterogeneous tissue straining and cyclic recruitment of lung units that close during each expiration~\cite{Kollisch2018}. Some computational studies have attempted to model the overall recruitment behavior of lungs and show a limited degree of predictive capability~\cite{Nabian2018,Morton2020,Zhou2021}. Due to their single-compartment design, however, they cannot inform about potential sites of regional overdistension caused by tissue inhomogeneity. Multi-compartment models, on the other hand, allow us to investigate the complex dynamics of lung recruitment and derecruitment at a finer level of spatial and temporal scale~\cite{Sundaresan2011,Sundaresan2009a,Bates2002,Ma2011}. In particular, the empirical model of time-dependent recruitment and derecruitment (R/D) dynamics introduced by Bates~et~al.~\cite{Bates2002,Massa2008,Ma2010} has been widely used to interpret experimental data from animal models~\cite{Ma2011,Massa2008,Smith2015,Broche2017,Knudsen2018,Smith2013a} and to link R/D dynamics to VILI in certain ventilation strategies~\cite{Ma2011,Ma2010,Smith2015,Broche2017,Knudsen2018,Smith2013a,Hamlington2016}. This model has not, however, been incorporated into an anatomically and physiologically realistic representation of the human lung, nor has it been personalized to represent the pathology of an individual patient. In contrast, we have developed a physics-based computational model of the lung~\cite{Roth2017a,Roth2017,Ismail2013} that is able to reproduce the pulmonary ventilation of an ARDS patient over both global and local length scales, but this model does not incorporate a representation of the dynamics of R/D that is so crucial to the fate of a lung with ARDS.

The goal of the present work is therefore to integrate the dynamics of R/D into a comprehensive and anatomically based computational lung model. To do this, we combine our physics-based reduced-dimensional model of the lung based on a realistic morphology~\cite{Roth2017a,Roth2017,Ismail2013} with the afore-mentioned empirical model of R/D dynamics~\cite{Bates2002,Massa2008,Ma2010,Smith2015,Smith2013a}. To enhance the physical foundation of the model, the R/D dynamics are related to the dimensions of the airways and to the biophysical properties of the airway lining fluid. This novel modeling approach allows a more realistic estimation of how high-stress sites within the lungs of a given patient might act as foci for the development and propagation of VILI. We evaluate our new model by matching its mechanical behavior to clinical data from a patient suffering from ARDS, and then we used the model to simulate several ventilation maneuvers undertaken at the bedside. Our goal is to develop a tool for creating personalized therapies for the mechanically ventilated patient.

\section{Materials and methods}
\label{sec:2}

\begin{figure*}[htb!] 
\begin{center}
 \includegraphics[width=1.0\textwidth]{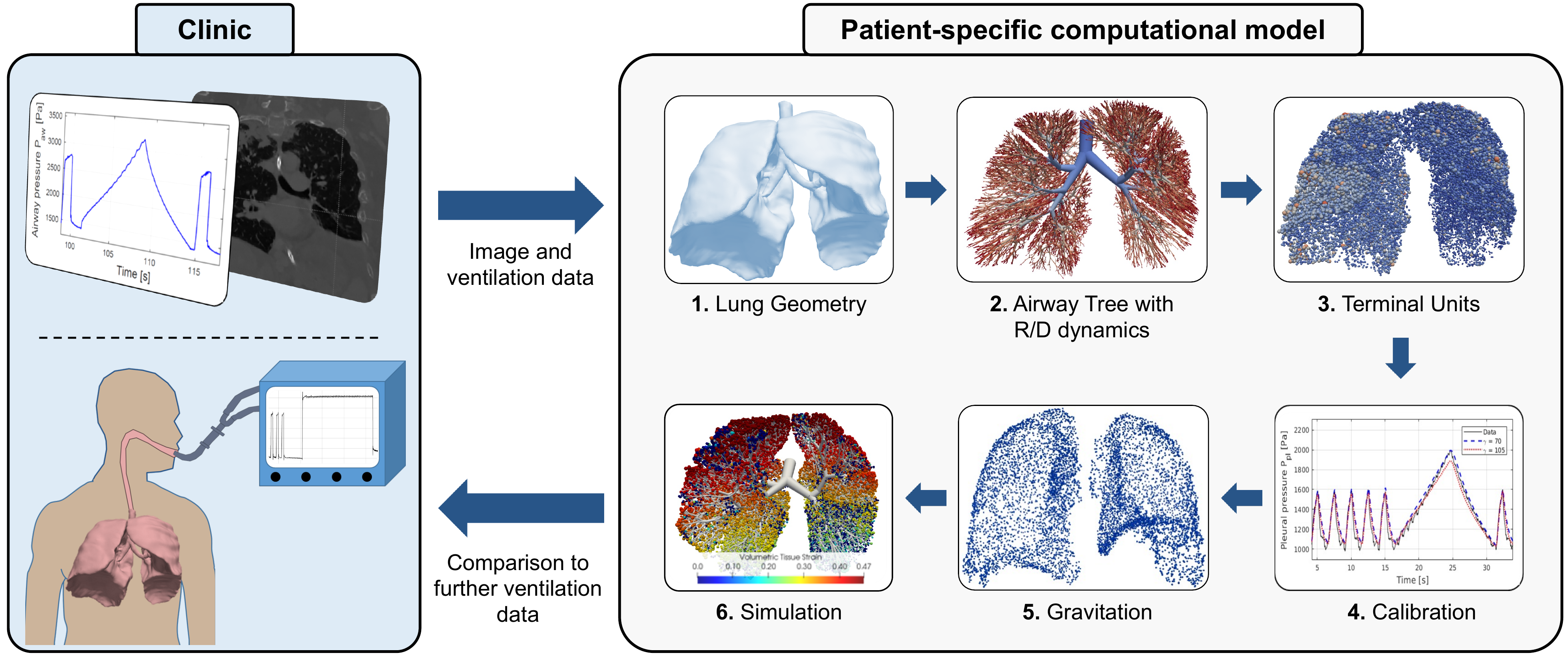} 
\end{center}  
\caption{Schematic overview of the process to generate the patient-specific computational model of the lung from clinical data to simulate various ventilation profiles and to retrospectively compare it to the clinical reference measures.}
\label{modelgeometry}       
\end{figure*}

To set up our patient-specific computational model, information about lung geometry and pulmonary pathophysiology is extracted from clinical data and we further used the data to calibrate the mechanical behavior of the model to that of the patient (see schematic outline in Figure~\ref{modelgeometry}).

\subsection{Clinical data}
\label{sec:2:data}

We used chest CT images and ventilation data acquired during the medical treatment of a critically ill, endotracheally intubated 50 year-old female patient suffering from ARDS who was included in another study~\cite{Becher2021g}. The data were provided in an anonymized format by the Department of Anesthesiology and Intensive Care Medicine at Christian Albrechts University in Kiel. Ethical approval was obtained from the ethics committee of the Medical Faculty in Kiel, and the underlying study was carried out in accordance with the Declaration of Helsinki. 

\begin{figure*}[htb!] 
\begin{center}
\includegraphics[width=0.4\textwidth]{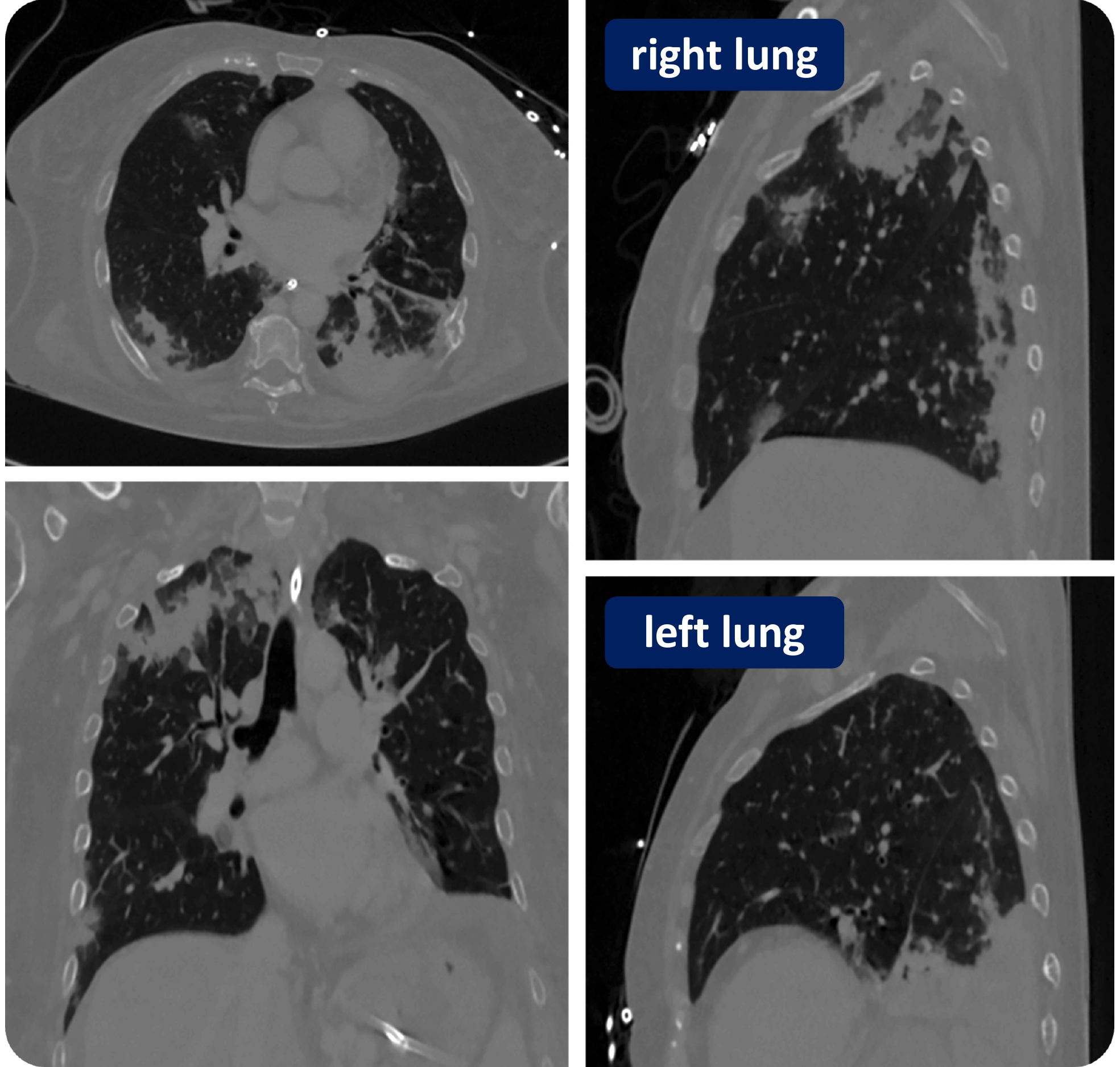} 
\end{center}  
\caption{CT scan of the patient in axial, coronal and sagittal (right and left lung) view exhibiting heterogeneous lung regions.}
\label{CT}       
\end{figure*}

\textbf{Image data}\quad A single three-dimensional thoracic CT scan of the patient provides the overall geometry for the lung model and allows us to identify derecruited, and thereby potentially recruitable, regions. The scan (512x512x1062 pixels each having dimensions 0.98x0.98x0.7mm) was recorded at a PEEP of 10~mbar (PEEP10). Exemplary views of the lung showing the heterogeneous injury of the organ are depicted in Figure~\ref{CT}.

\textbf{Ventilation data}\quad Various ventilation profiles were applied to the patient at the bedside to reveal the specific mechanical properties of the respiratory system and to adapt the ventilation parameters appropriately for therapy. Measurements included the pressure at the airway opening, the tracheal airflow entering the lung, and esophageal pressure as a surrogate for pleural pressure. Transpulmonary pressure~($P_{\mathrm{tp}}$) was calculated as the difference between tracheal and esophageal pressures.

\subsection{Reduced-dimensional lung model}
\label{sec:2:model}

Our computational study is based on the previously developed reduced dimensional computational model~\cite{Roth2017a,Roth2017,Ismail2013} extended to include a well-investigated model of R/D dynamics \cite{Bates2002,Massa2008,Ma2010}. In the following, we briefly restate the central components of these models.

\subsubsection{Model geometry}
\label{sec:2:modelgeneration}

Using the CT image, we identified the centerline of the visible parts of the airway tree from distal end of the endotracheal tube down to the lobar bronchi. The individual lobes of the lung were segmented (Mimics and 3-Matic, Materialise, Leuve, Belgium) and the airway branches within each lobe generated by a space-filling algorithm~\cite{Ismail2013,HowatsonTawhai2000}. The recursive branching of a parent airway into two daughter airways follows morphological length and diameter ratios~\cite{Weibel1963,Horsfield1971,Majumdar2005} and terminates when either
\begin{itemize}
    \item the length of an airway is smaller than 1.2~$\mathrm{mm}$,
    \item the diameter of an airway is smaller than 0.4~$\mathrm{mm}$,
    \item a maximal number of 17 generations is reached, or
    \item the hull geometry of the segmented lobe is penetrated.
\end{itemize}
The resulting three-dimensional airway tree mimics the purely conducting zone of the lung and is modeled by reduced-dimensional airway elements (see Section~\ref{sec:2:modelairways}). At each terminal branch, we attach a so-called \textit{terminal unit} (see Section~\ref{sec:2:modelterminalunits}). These terminal units represent the remaining smaller tissue structures reaching into the parenchymal region beyond the conducting airway tree. In total, the model contains 48,178 reduced airway elements and 24,089 terminal units.

\subsubsection{Conducting airways}
\label{sec:2:modelairways}

\textbf{0D airway model}\quad Each branch of the tracheo-bronchial tree is modeled by a reduced-dimensional element reproducing the averaged behavior of flow and wall mechanics in a fully resolved, elastic, three-dimensional airway. \textcolor{black}{The equations underlying a 0D airway are derived from the Navier Stokes equations expressing conservation of mass~(resulting in Equ.~\eqref{eq:0D_pipe1}) and balance of momentum for Newtonian fluids~(resulting in Equ.~\eqref{eq:0D_pipe2}), and include the mechanics of the airway wall by a relationship between pressure and cross-sectional area. We followed closely the symmetric 0D model presented in~\cite{Formaggia2009} for the circulatory system, but extended it for airways according to~\cite{Ismail2013}. The main assumptions underlying the derivations are axial symmetry, small curvature, and constant geometric and material parameters, e.g., cross-sectional area, along the longitudinal axis of each airway element.} The rates of inflow~$Q_{\mathrm{in}}$ and outflow~$Q_{\mathrm{out}}$ of a 0D airway are driven by the pressure drop $\Delta P = P_{\mathrm{in}} - P_{\mathrm{out}}$ across the element and the external pressure,~$\widetilde{P}_{\mathrm{ext}}$, and are determined by
%
    \begin{flalign}
	C \dfrac{\mathrm{d}}{\mathrm{d}t}\left( \dfrac{1}{2} \left( P_{\mathrm{in}} + P_{\mathrm{out}} \right) - \widetilde{P}_{\mathrm{ext}}\right)  + Q_{\mathrm{out}} - Q_{\mathrm{in}} + C\cdot R_{\mathrm{visc}} \dfrac{\mathrm{d}}{\mathrm{d}t} \left( Q_{\mathrm{out}} - Q_{\mathrm{in}}\right) = 0, \label{eq:0D_pipe1} \\
    \dfrac{I}{2} \dfrac{\mathrm{d}}{\mathrm{d}t} \left(Q_{\mathrm{in}} + Q_{\mathrm{out}}\right) + \dfrac{1}{2} \left( R_{\mathrm{\mu}} + R_{\mathrm{conv}}\right) \cdot \textcolor{black}{\left( Q_{\mathrm{in}} + Q_{\mathrm{out}} \right)}  + P_{\mathrm{out}} - P_{\mathrm{in}} = 0, \label{eq:0D_pipe2}
    \end{flalign}
where $C$ is the capacitance of the airway wall, $R_{\mathrm{visc}}$, $R_{\mathrm{conv}}$ and $R_{\mathrm{\mu}}$ are the visco-elastic, convective and nonlinear airway resistances, respectively, and $I$ is inductance. \textcolor{black}{See Appendix~\ref{sec:app:airways} for more details on these quantities and model assumptions.}
\textcolor{black}{$\widetilde{P}_{\mathrm{ext}}$ is the pressure of neighbouring lung regions acting on a reduced airway element, i.e.,~the internal pressure of the terminal unit closest to the longitudinal axis of the specific airway.}

\textbf{R/D dynamics of airways} \quad To include time-dependent R/D dynamics in our patient-specific model of the lung, we followed the semi-empirical and established approach introduced by~\cite{Bates2002,Massa2008,Ma2010}. The approach assumes that an individual (de-)recruitable airway can either be completely open (state = 1) or completely closed (state = 0), and switches between these states depending on the pressure it is exposed to and the duration of this exposure. The transitions between the two states are modeled using a variable~$x$ that moves along a virtual trajectory between 0 and 1 according to
\begin{equation}
\frac{\mathrm{d}x}{\mathrm{d}t} = 
\left\{
\begin{aligned}
s_{\mathrm{o}}(P_{\mathrm{in}} - P_{\mathrm{o}}) \qquad P_{\mathrm{in}} > P_{\mathrm{o}}\\
s_{\mathrm{c}}(P_{\mathrm{in}} - P_{\mathrm{c}}) \qquad P_{\mathrm{in}} < P_{\mathrm{c}}\\
0 \qquad  else,
\label{eq:reopening}
\end{aligned}
\right.
\end{equation}
where $P_{\mathrm{o}}$ is the critical opening pressure, $P_{\mathrm{c}}$ is the critical closing pressure, and $s_{\mathrm{o}}$ and $s_{\mathrm{c}}$ are constants of proportionality. When $P_{\mathrm{in}}$ exceeds $P_{\mathrm{o}}$, $x$ moves toward 1. If $P_{\mathrm{in}}$ falls below $P_{\mathrm{c}}$, $x$ approaches 0. The rate at which $x$ changes depends both on the constants $s_{\mathrm{o}}$ and $s_{\mathrm{c}}$ and on the difference between $P_{\mathrm{in}}$ and the corresponding critical pressure. If $x$ reaches 0 when the airway is open, or 1 when the airway is closed, closure or opening is triggered, respectively. When 0~$ < x < $~1, the state of an airway does not change. When $P_{\mathrm{c}} < P_{\mathrm{in}} < P_{\mathrm{o}}$, $x$ remains constant.
Following~\cite{Bates2002}, we model a closed 0D airway element by setting its resistance to a high value of $R_{\mathrm{\mu}} = 10^{16} \mathrm{kg/s.m^4}$ in Equ.~\eqref{eq:0D_pipe2}, which effectively eliminates airflow through it. Further details on the model can be found in~\cite{Bates2002,Massa2008,Ma2010}. 

In previous works employing this dynamic R/D model, the values of the model parameters in Equ.~\eqref{eq:reopening} have generally \textcolor{black}{been} determined in a stochastic manner~\cite{Bates2002,Massa2008,Ma2010}, which represents the complex mechanical events occurring in an airway during reopening in purely phenomenological terms. In contrast, we try to base the parameters of our composite model on physical relationships as far as possible. Thus, based on experimental findings~\cite{Naureckas1994}, we compute $P_{\mathrm{o}}$ for an airway using its radius~$r_{\mathrm{aw}}$ \textcolor{black}{resulting from the geometry generation} and the surface tension,~$\gamma$, of its lining fluid according to
\begin{align}
	P_{\mathrm{o}} = 8.3 \frac{\gamma}{r_{\mathrm{aw}}}.
	\label{eq:reopPressure}
\end{align}
Since the composition of the lining fluid in a diseased lung region is unknown, we consider three different values for~$\gamma$ in Equ.~\eqref{eq:reopPressure}. As one reasonable choice, we assume the fluid to be dominated by human serum albumin, a major component of human blood plasma, with a surface tension of~$\gamma = 70\ \text{dyn/cm}$~\cite{VanOss1981}. This protein is presumably also present in airways damaged by inflammatory processes and cyclic reopening~\cite{Bilek2003}, resulting in epithelial cell damage and leakage of blood plasma into the air spaces. As the airway fluid might also transition to stickier sputum~\cite{Widdicombe2002j}, we consider two additional arbitrarily chosen scenarios of $\gamma = 100\ \text{dyn/cm}$ and $\gamma = 130\ \text{dyn/cm}$. 
\textcolor{black}{Note that Equ.~\eqref{eq:reopPressure} may no longer apply to mucus due to the non-Newtonian properties of the fluid. However, there is no known relationship between the critical opening pressure (or opening velocity) in the airways and a non-Newtonian fluid. To avoid introducing further uncertainties due to arbitrary mathematical relationships and unknown variables, we use the present relationship as a preliminary solution to obtain the higher critical opening pressures of airways that are likely to occur in airways when exposed to more severe pathology and filled with a sticky fluid.}

We choose $P_{\mathrm{c}}$ in Equ.~\eqref{eq:reopening} 4\,$\mathrm{cmH_2O}$ lower than the corresponding $P_{\mathrm{o}}$ according to~\cite{Ma2010}. The constants $s_{\mathrm{o}}$ and $s_{\mathrm{c}}$ that influence the rate of airway opening and closing, respectively, are assumed to follow the quasi-hyperbolical distributions~$s_{\mathrm{o}} \in\frac{S_{\mathrm{o}}}{\mathrm{unif}[0,1]}$ and $s_{\mathrm{c}} \in\frac{S_{\mathrm{c}}}{\mathrm{unif}[0,1]}$~\cite{Ma2010}. Herein, $\mathrm{unif}[0,1]$ describes uniformly distributed stochastic values between 0 and 1. $S_{\mathrm{o}}$ and $S_{\mathrm{c}}$ are constants set to 0.04\,$\mathrm{cmH_2O}^{-1} \mathrm{s}^{-1}$ and 0.004\,$\mathrm{cmH_2O}^{-1} \mathrm{s}^{-1}$, respectively~\cite{Ma2010}. Finally, $s_{\mathrm{o}}$ and $s_{\mathrm{c}}$ are coupled such that $s_{\mathrm{o}} = 10s_{\mathrm{c}}$ for each collapsible airway.

\begin{figure}[htbp]
	\centering
	\includegraphics[width=0.4\linewidth]{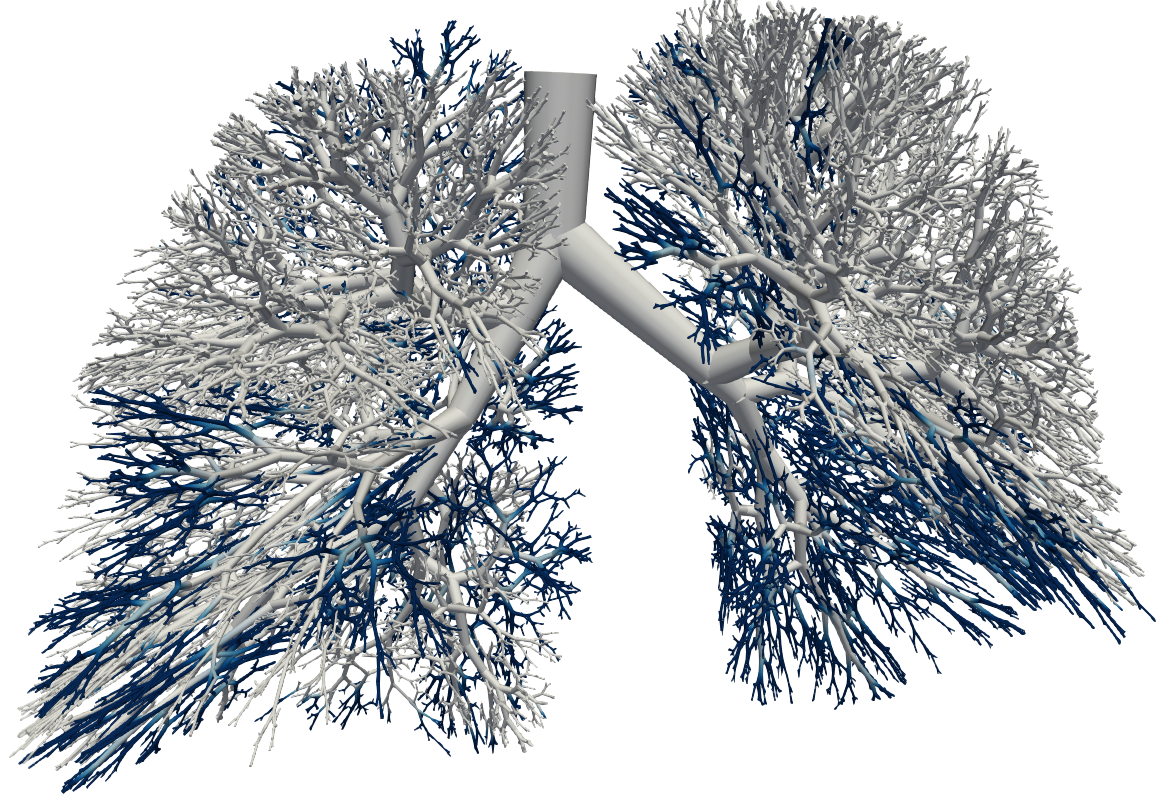}
	\caption[Generated airway tree with (de-)recruitable airways (black)]{Generated airway tree with (de-)recruitable airways (black) identified by regions in CT scan with densities above -300 HU.}
	\label{fig:airways_derecruit}
\end{figure}

\begin{figure}[htbp]
	\centering
	\includegraphics[width=0.6\linewidth]{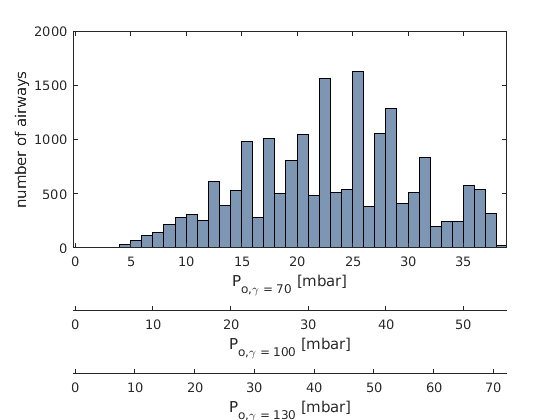}
	\caption[Distribution of critical opening pressures]{Distribution of the critical opening pressures $P_{\mathrm{o, \gamma}}$ of all recruitable airways with respect to the three investigated values for the surface tension $\gamma$, 70\,dyn/cm, 100\,dyn/cm and 130\,dyn/cm.}
	\label{fig:histogram_p_open}
\end{figure}

To account for the patient-specific spatial heterogeneity of the diseased lung in our computational model, we apply the above time-dependent R/D model to those airways of the tracheo-bronchial tree that are located in injured (i.e., non- or poorly-aerated) regions of the lung. The potentially atelectatic regions are identified in the CT scan as having Hounsfield Units~(HU)~$> -300$~\cite{Markstaller2003}. All airways distal to a collapsible airway are assumed to be of a similarly diseased state (e.g., due to abnormal liquid lining properties and/or restricted airflow) and are therefore also subjected to R/D in our model. Figure~\ref{fig:airways_derecruit} shows all 18,952 airways of the airway tree that are collapsible based on the conditions described above.
The distributions of $P_{\mathrm{o}}$ for all collapsible airways, following Equ.~\eqref{eq:reopPressure}, are depicted in Figure~\ref{fig:histogram_p_open} for each~$\gamma$.

\subsubsection{0D terminal units}
\label{sec:2:modelterminalunits}

The terminal units comprising the respiratory zone beyond the distal ends of the conducting airway tree are represented by generalized Maxwell models that reproduce the nonlinear viscoelastic behavior of alveolar tissue~\cite{Ismail2013}. A hyperelastic compressible Ogden-type material law~\cite{Ogden1972} is used to model the nonlinear elastic behavior of the lung parenchyma~\cite{Roth2017a}, derived by assuming pure volumetric deformation as 
\begin{align}
P_{\mathrm{alv}} - P_{\mathrm{pl}} = \frac{\kappa}{\beta} \cdot \frac{V_{0,\mathrm{unit}}}{V_{\mathrm{unit}}} \left(1 -  \left( \frac{V_{0,\mathrm{unit}}}{V_{\mathrm{unit}}} \right)^{\beta} \right),
\label{eq:OgdenMaterial}
\end{align}
where $V_{\mathrm{unit}}$ is the current volume of a terminal unit, $V_{0,\mathrm{unit}}$ is the reference value of $V_{\mathrm{unit}}$ in the stress-free state, $P_{\mathrm{alv}}$ is alveolar pressure, and $P_{\mathrm{pl}}$ is pleural pressure. The slope and shape determining parameters~$\kappa$ and $\beta$, respectively, define the material properties of the lung tissue.
$\kappa$ and $\beta$ are estimated for the lungs of a specific patient by a fitting procedure~(Section~\ref{sec:2:modelfitting}). 

When determining $V_{0,\mathrm{unit}}$ for each terminal unit, we have to consider the heterogeneous nature of a damaged lung. The terminal units are mechanically independent and thus generally exist in different mechanical states, either collapsed and presumably stress-free or normally inflated and thus distended. We cannot therefore simply calculate the total lung volume from the CT image by solving Equ.~\eqref{eq:OgdenMaterial} and distributing it evenly across all terminal units because this could underestimate local increase in volume capacity resulting from recruitment. We therefore split the total volume of each lobe, derived from the CT image, into two parts, one consisting of air and the other of tissue based on HU~=~-1000 for air and HU~=~0 for water ($\sim$~tissue)~\cite{Gattinoni2001}. 

The volume of the part comprised of water is assumed to correspond to the volume of the lobar tissue, $V^{\mathrm{tissue,lb}}$. We distribute $V^{\mathrm{tissue,lb}}$ across all lobar terminal units in proportion to the area, $A_{\mathrm{unit}}^{\mathrm{lb}}$, of the airway supplying each unit, which was set during the tree growing procedure. $A^{\mathrm{lb}}$ is the sum of all the individual $A_{\mathrm{unit}}^{\mathrm{lb}}$ for the lobe. That is, $V^{tissue}_{\mathrm{unit}} = V^{\mathrm{tissue,lb}} \cdot A_{\mathrm{unit}}^{\mathrm{lb}} \slash A^{\mathrm{lb}}$. This approach is motivated by the assumption that larger airways support larger alveolar regions.
To get the total volume of a terminal unit, including both its air and tissue volumes, we distribute the total air volume of the lung across all terminal units so as to recreate their respective HU in the CT image. The total volume of a terminal unit determined in this way corresponds to the volume~$V_{\mathrm{unit}, CT}$ of the unit under the pressure load at the time of the CT scan, i.e., with PEEP10 applied at the airway opening and the corresponding pleural pressure. The reference volume~$V_{0,\mathrm{unit}}$ of open terminal units can then be calculated by solving Equ.~\eqref{eq:OgdenMaterial}. Initially trapped terminal units, i.e., units that are attached to an initially collapsed airway, are assumed to be nearly collapsed, and therefore already in a stress-free state so that $V_{0,\mathrm{unit}} = V_{\mathrm{unit}, CT}$.

\subsubsection{Pressure boundary conditions}
\label{sec:2:modelpleural}

The external pressure acting on a terminal unit has two components: 1) a volume-dependent component~$P_{\mathrm{pl}}^{\mathrm{vol}}$ due to the elastic recoil of the chest wall, and 2) a static component~$P_{\mathrm{pl}}^{\mathrm{weight}}$ due to the weight of the lung that is above the units in question. The pleural pressure is the sum of these two components, that is: \textcolor{black}{$P_{\mathrm{pl}} = P_{\mathrm{pl}}^{\mathrm{vol}} + P_{\mathrm{pl}}^{\mathrm{weight}}$.}

\textcolor{black}{$P_{\mathrm{pl}}^{\mathrm{vol}}$} is determined by the total volume of all terminal units and thus implicitly integrates the passive mechanical properties of the sedated chest wall into the model. Since the relationship between the lung volume and $P_{\mathrm{pl}}$ during a quasi-static inflation maneuver is linear over most of its typical range (Figure~\ref{fig:fit}b, see more details in Section~\ref{sec:2:modelfitting}), we match the linear relationship
\begin{align}
P_{\mathrm{pl}}^{\mathrm{vol}} = P_{\mathrm{pl,0}} + P_{\mathrm{pl,lin}} V_{\mathrm{frac}},
\label{eq:pleural_pressure_volume_dep} 
\end{align}
with 
\begin{align}
V_{\mathrm{frac}} = \frac{(V - V_{\mathrm{PEEP}})}{(V_{\mathrm{max}} - V_{\mathrm{PEEP}})} \:
\end{align}
to clinical measurements by determining $P_{\mathrm{pl}, 0}$ and $P_{\mathrm{pl,lin}}$ accordingly. The volume fraction~$V_{\mathrm{frac}}$ describes the ratio between the increase in volume, $V - V_{\mathrm{PEEP}}$, from the volume level at PEEP10, calculated from the CT scan, and the increase in volume during the quasi-static inflation maneuver at end-inspiration, $V_{\mathrm{max}} - V_{\mathrm{PEEP}}$.

$P_{\mathrm{pl}}^{\mathrm{weight}}$ is the pressure across a section of lung tissue due to the weight of the lung above it, treating the organ as a fluid body subjected to gravity. This weight is usually larger in an ARDS compared to a healthy lung because of the extra fluid accumulation. To include this effect in our model, we employ the relation proposed in~\cite{Pelosi1994} that provides the pressure resulting from the weight of the lung as a function of ventral-to-dorsal height of the lung, $h_{\mathrm{lung}}$:
\begin{align}
P_{\mathrm{pl}}^{\mathrm{weight}} = 0.541 \cdot (h_{\mathrm{lung}} - h_{\mathrm{balloon}}) + 0.015 \cdot (h_{\mathrm{lung}}^{2} - h_{\mathrm{balloon}}^{2}).
\label{eq:pressure_gradient_superimposed} 
\end{align}
To achieve a static pressure of zero at the reference point of measurement, which is made using an esophageal balloon, we determine the height of the measurement site, $h_{\mathrm{balloon}}$, from the CT image~\cite{Yoshida2018j} and introduce it into Equ.~\eqref{eq:pressure_gradient_superimposed}.

\subsubsection{Patient-specific parameter calibration}
\label{sec:2:modelfitting}

\begin{figure}[htbp]
	\centering
    \includegraphics[trim=0 0 0 0,clip,width=1.0\textwidth]{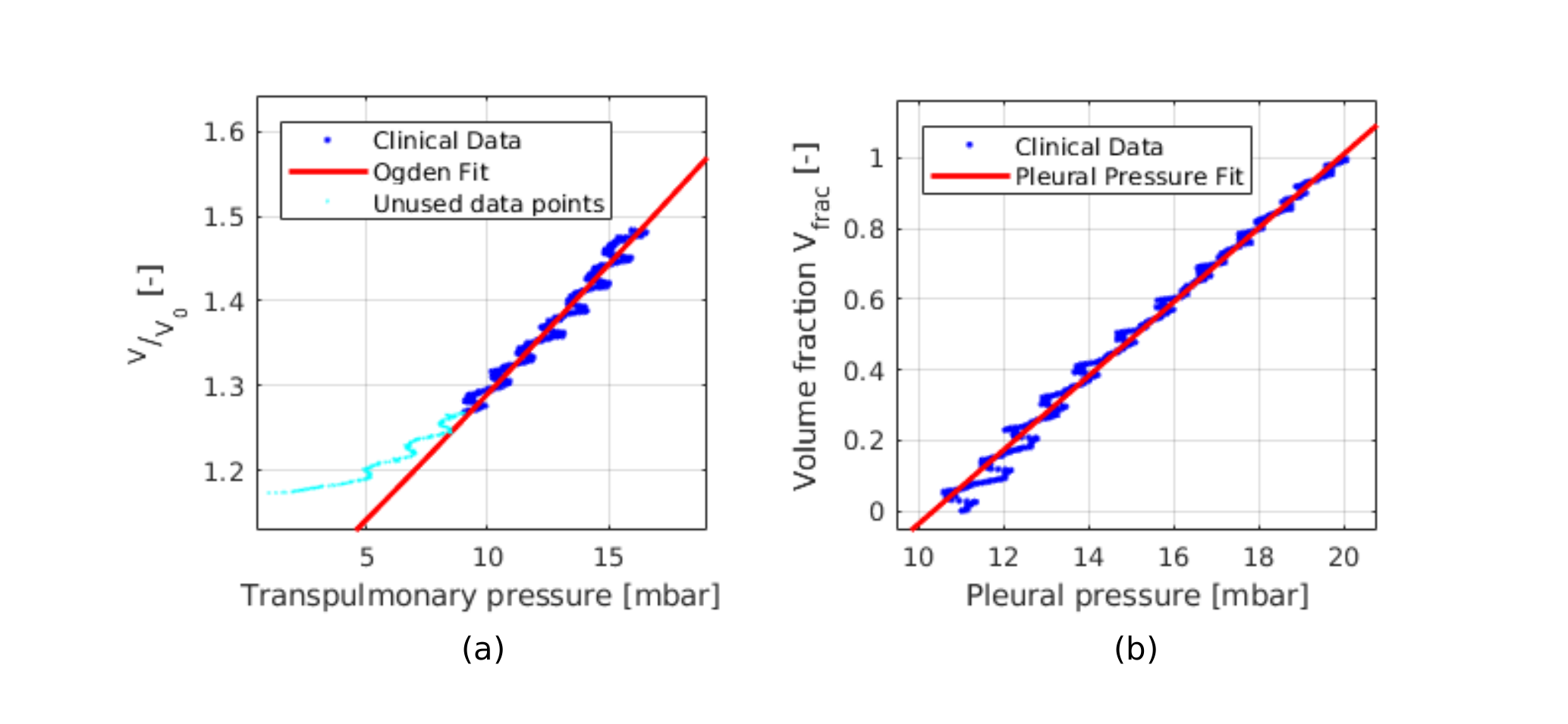}
	\caption[Parameter calibration]{(a) Pressure-volume curve of all terminal units (red) calibrated to the clinical measurement of transpulmonary pressure and volume (blue), and (b) relation between the volume fraction and the pleural pressure used as boundary condition in the model (red) determined from the measured pleural pressure-volume-curve (blue). Data points in (a) below the lower inflection point (light blue) have been neglected for the calibration as the curvature of this part might result from recruitment effects and, thus, not exhibit the pure elastic pressure-volume behavior of the lung.}
	\label{fig:fit}
\end{figure}

At an overall level, the respiratory system is composed of two distinct interacting sub-systems, the lung and the chest wall~\cite{West2016}. In order to personalize these two sub-systems to the patient in this study, we fit the remaining parameters of Equ.~\eqref{eq:OgdenMaterial} (i.e., $\kappa$ and $\beta$) and Equ.~\eqref{eq:pleural_pressure_volume_dep} (i.e., $P_{\mathrm{pl,0}}$ and $P_{\mathrm{pl,lin}}$) to recordings of pressures and volume made during a quasi-static inflation maneuver starting at PEEP10. The low flow employed during this maneuver minimizes viscous effects in the lung tissue, so the resulting pressure-volume curve is largely reflective only of the elastic behavior of the respiratory system. Both equations were fit to the inspiratory segments of the pressure-volume data in MATLAB using a nonlinear regression method (see Figure~\ref{fig:fit}). 

In our model, the lung sub-system can be viewed as the collective behavior of all terminal units (see Section~\ref{sec:2:modelterminalunits}). As their deformation is governed by the interplay of the alveolar and pleural pressures, we fit $\kappa$ and $\beta$ to the transpulmonary pressure $P_{\mathrm{tp}} = P_{\mathrm{alv}} - P_{\mathrm{pl}}$ and the corresponding lung volume in the quasi-static inflation maneuver (Figure~\ref{fig:fit}a). We assume $\kappa$ and $\beta$ to be the same for all terminal units. The heterogeneity of the diseased lung is accounted for by collapsible airways. That is, when an airway is closed its downstream tissue units no longer communicate with the airway opening (i.e., these units become derecruited). As a consequence, the model stiffens regionally. 

Accordingly, we estimate the parameters of $P_{\mathrm{pl}}^{\mathrm{vol}}$ (Equ.~\eqref{eq:pleural_pressure_volume_dep}) such that we can reproduce the $P_{\mathrm{pl}}$-$V$ behavior of the quasi-static inflation maneuver (Figure~\ref{fig:fit}b). 
The values of the parameters describing the elastic material behavior of the terminal units and the pleural pressure conditions are given in Table~\ref{tab:fit_parameters}.

\section{Results}
\label{sec:results}

\begin{figure}[htbp!]
	\centering
        \includegraphics[trim=0 0 0     0,clip,width=1.0\textwidth]{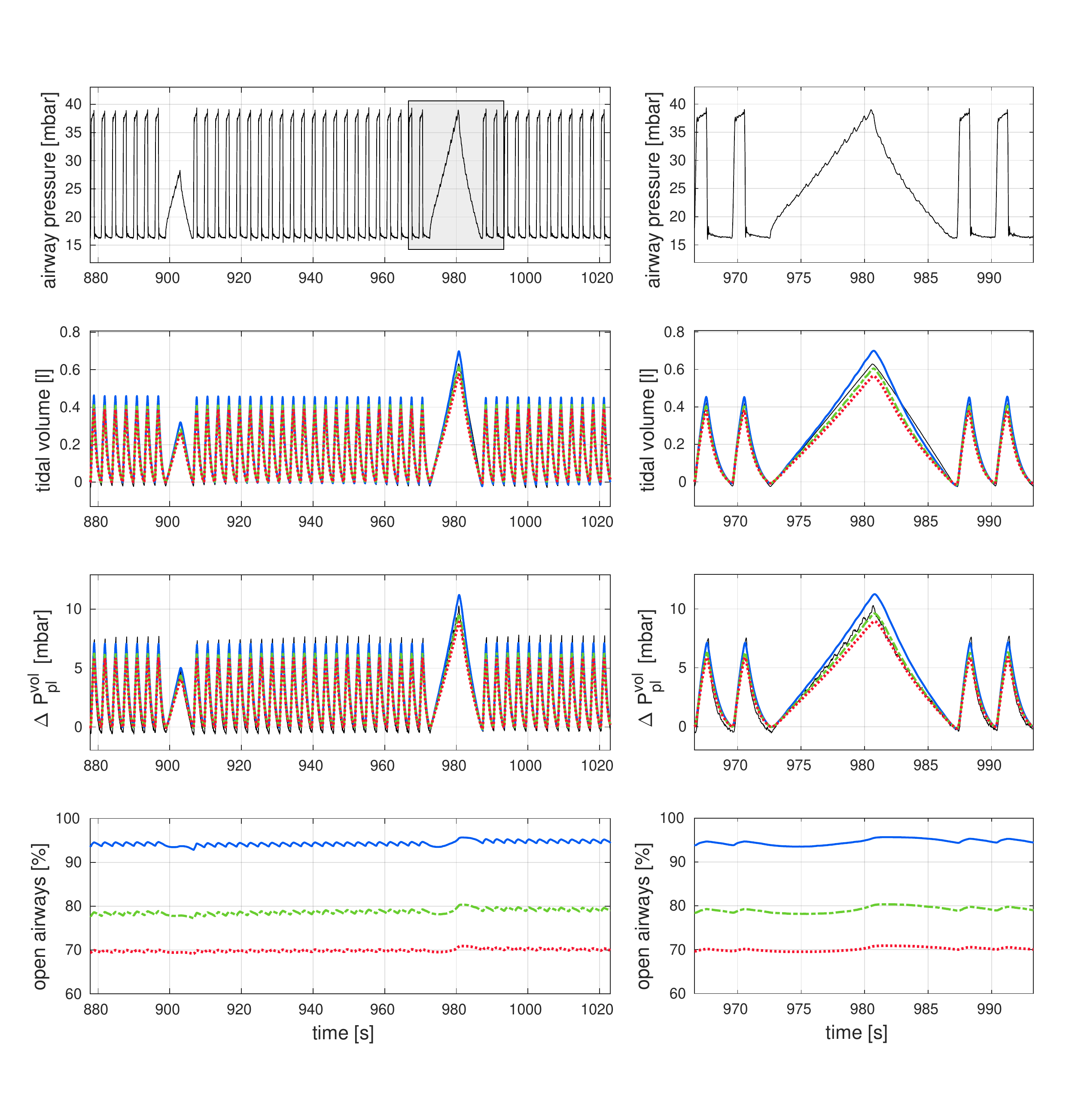}
	    \caption{Simulation results of the proposed computational model for a clinically applied airway pressure profile of normal ventilation alternated by two quasi-static low-flow inflation maneuvers (top) for all three scenarios of surface tension~$\gamma$, 70~dyn/cm (blue solid), 100~dyn/cm (green dashed) and 130~dyn/cm (red dotted) from top to bottom: tidal volume and change in pleural pressure compared to the clinical measurements (black solid), and the percentage of open airways in the whole model; in each figure, the area highlighted in grey (top left) indicates the period shown in more detail in the right column for all mentioned measures.}
	    \label{fig:results_lf}
\end{figure}

\begin{figure}	
	\centering
	    \includegraphics[trim=0 0 0 0,clip,width=1.0\textwidth]{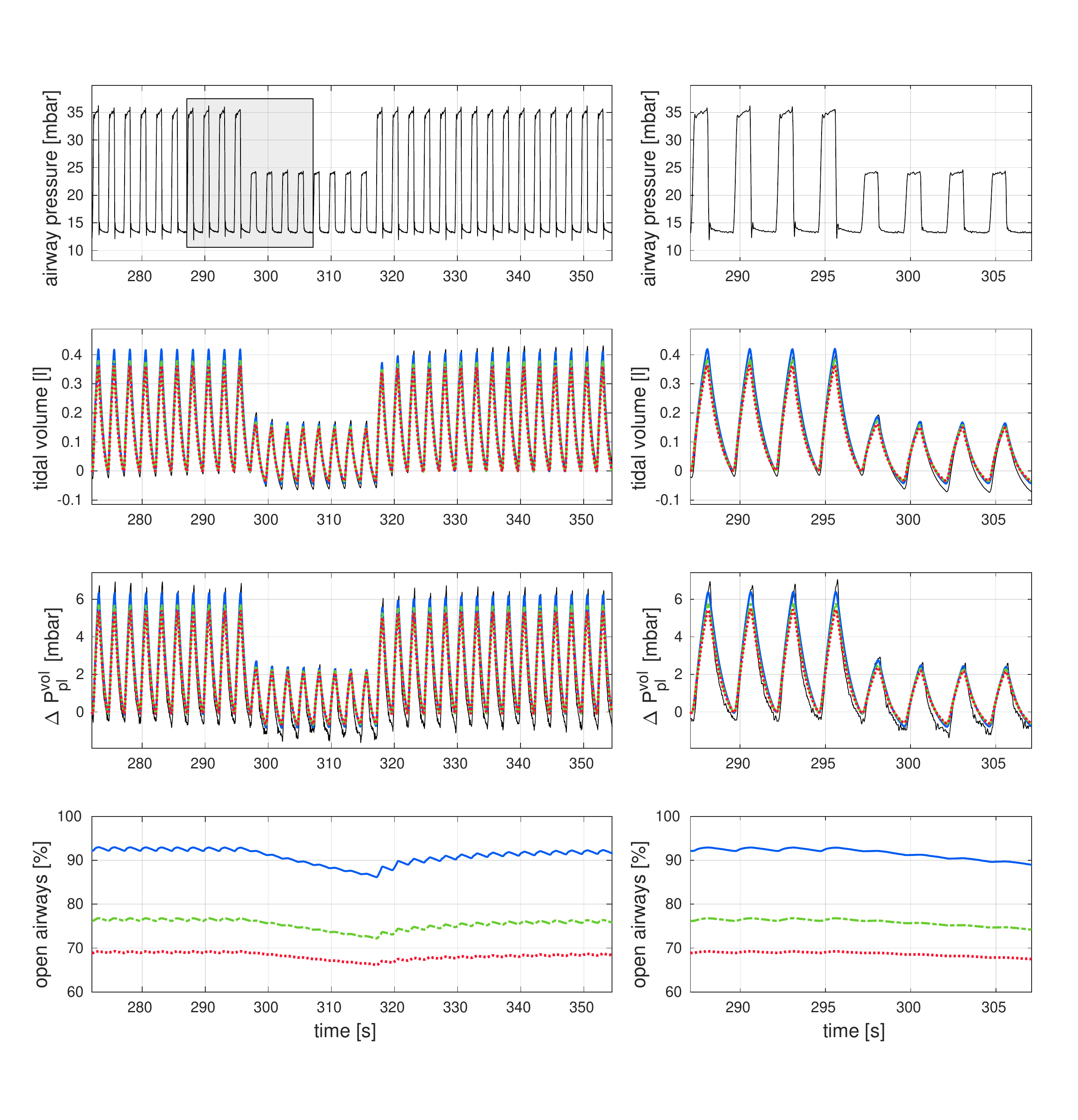}
	    \caption{Simulation results of the proposed computational model for a clinically applied airway pressure profile of normal ventilation with temporarily halved driving pressure (top) for all three scenarios of surface tension~$\gamma$, 70~dyn/cm (blue solid), 100~dyn/cm (green dashed) and 130~dyn/cm (red dotted) from top to bottom: tidal volume and change in pleural pressure compared to the clinical measurements (black solid), and the percentage of open airways in the whole model; in each figure, the area highlighted in grey (top left) indicates the period shown in more detail in the right column for all mentioned measures.}
	    \label{fig:results_deltap1}
\end{figure}

\noindent\textbf{Simulation protocol} \quad To investigate the behavior of our model in a real-life scenario, we simulated a 28-minute long recording of the patient's ventilation protocol at the bedside. The model was driven by the airway pressure applied during mechanical ventilation. Note that the simulated ventilation patterns do not include the quasi-static inflation maneuver from PEEP10 that was used to calibrate the model (Section~\ref{sec:2:modelfitting}), so the maneuvers used to validate the model are not the same as that used to develop it. \\
We used the following maneuvers for validation:
\begin{enumerate}
    \item cycles of normal ventilation containing two quasi-static inflation maneuvers with different peak pressures, starting from PEEP16 (Figure~\ref{fig:results_lf}), 
    \item cycles of normal ventilation at PEEP13 with half the original driving pressure (Figure~\ref{fig:results_deltap1}), and
    \item a sustained-inflation maneuver starting at PEEP16 in which the airway pressure was maintained at 40~mbar for a period of $\sim$32 seconds, followed by ventilation at elevated PEEP19.
\end{enumerate}
The chronological order of these maneuvers is indicated by the time indications in the figures. We simulated these maneuvers for three different values of $\gamma$ of 70, 100, and 130~dyn/cm. As we did not know the initial state of each collapsible airway, we chose $P_{o} > $~24~mbar as the critical threshold for airways to be declared closed initially. This value minimizes the transient opening and closing of airways during the first simulated ventilation cycles with PEEP10 and an end-inspiratory pressure of 32~mbar. We simulated several additional normal ventilation cycles at the beginning of each run to achieve steady state.\\

\begin{figure}
	\centering
	    \includegraphics[trim=0 0 0 0,clip,width=1.0\textwidth]{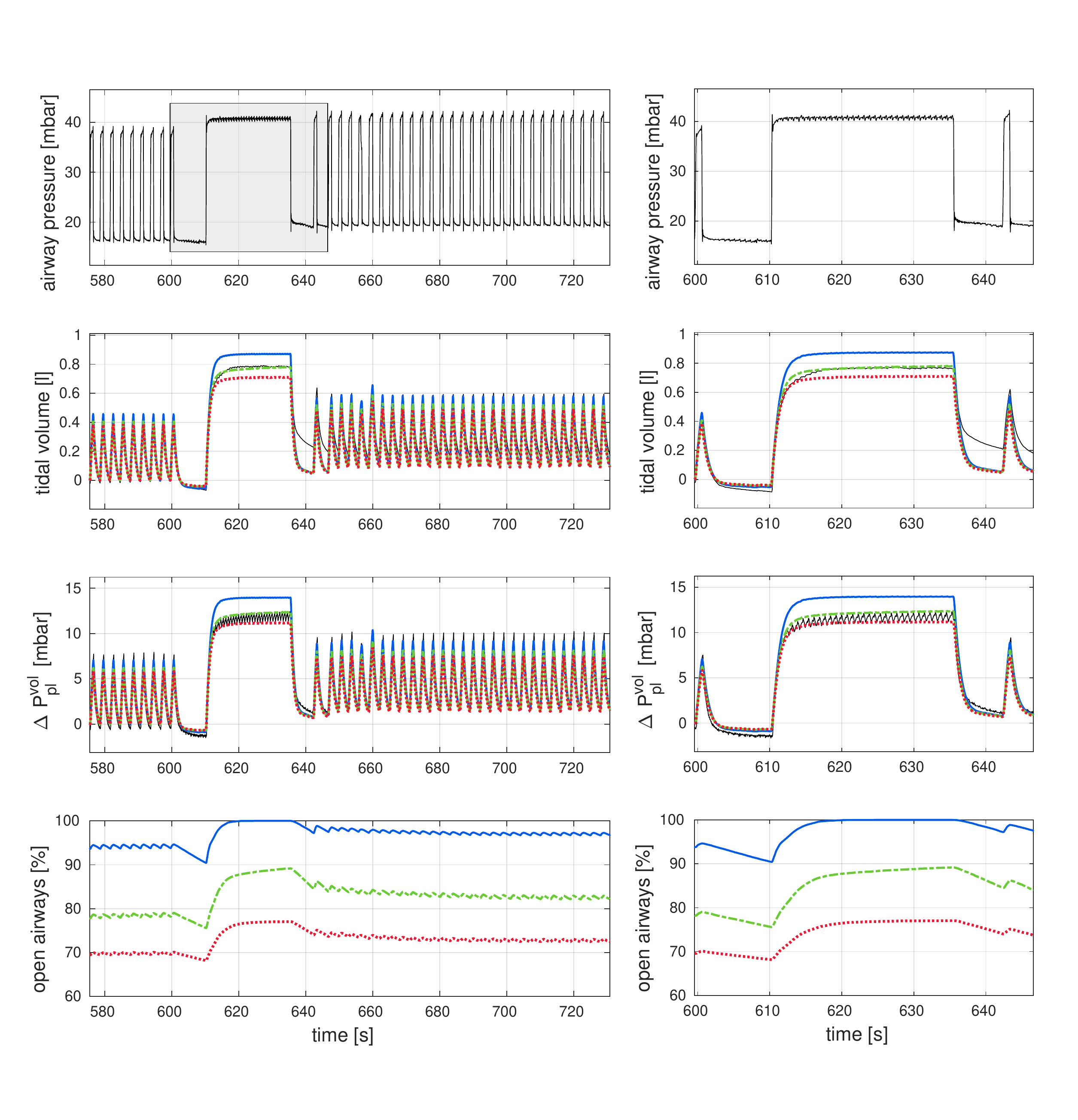}
	    \caption{Simulation results of the proposed computational model for a clinically applied airway pressure profile containing a sustained-inflation recruitment maneuver (top) for all three scenarios of surface tension~$\gamma$, 70~dyn/cm (blue solid), 100~dyn/cm (green dashed) and 130~dyn/cm (red dotted) from top to bottom: tidal volume and change in pleural pressure compared to the clinical measurements (black solid), and the percentage of open airways in the whole model; in each figure, the area highlighted in grey (top left) indicates the period shown in more detail in the right column for all mentioned measures.}
    	\label{fig:results_rm3}
\end{figure}

\noindent\textbf{Global ventilation quantities} \quad Figures~\ref{fig:results_lf}~-~\ref{fig:results_rm3} show the responses of relevant ventilation characteristics to the applied airway pressure profiles (top) over time. These include (from top to bottom): (i) airflow-derived tidal volume, (ii) comparison of measured and simulated $P_{\mathrm{pl}}$, and (iii) the percentage of open airways. Our simulated results show good agreement with the clinical measurements for all chosen values of $\gamma$. Nevertheless, as expected, the system is sensitive to the choice of~$\gamma$. In particular, a higher value of $\gamma$ leads to elevated and more broadly distributed values for $P_{\mathrm{o}}$ (Figure~\ref{fig:histogram_p_open}), which results in reduced tidal volumes and reduced swings in $P_{\mathrm{pl}}$ due to fewer reopened tissue regions (terminal units). We observe this effect in all investigated ventilation profiles at end-inspiration~(Figures~\ref{fig:results_lf}~-~\ref{fig:results_rm3}). 

For the normal ventilation cycles (Figure~\ref{fig:results_lf}) and the quasi-static inflation maneuver, the clinical data are best reproduced by the model for $\gamma = 100\ \text{dyn/cm}$, particularly as regards tidal volume. Assigning $\gamma$ the value for human albumin of $70\ \text{dyn/cm}$ leads to slight overestimations of tidal volume and swings in pleural pressure. When the driving pressure is reduced by 50\%, the tidal volume approximates clinical measurements at the peak of inspiration well~(Figure~\ref{fig:results_deltap1}). Nevertheless, there remains some mismatch between measured and simulated tidal volumes for all values of~$\gamma$ investigated. 
The simulated and measured sustained-inflation maneuvers show good agreement both in terms of tidal volume and change in pleural pressure for $\gamma = 100\ \text{dyn/cm}$, including their time-dependent, continuous increase (Figure~\ref{fig:results_rm3}). However, the simulated drop in pleural pressure and especially the observed simultaneous decrease in lung volume after this maneuver diverge from the clinical data. Elevations in both quantities are expected because of the raising of PEEP from 16~to~19~mbar, which would increase the volume capacity of the lungs due to both stronger distension and recruitment. The discrepancies between the simulations and the data could thus indicate that airways close too slowly in our model. This would allow air to escape from the lungs prior to closure and thus to reduce gas trapping. In addition, the differences between simulation and data differ for tidal volume and pleural pressure after the recruitment maneuver. This might suggest erroneous measurement of the very high flows occurring just after release of the high pressure, something that we also observed subsequent to other simulated recruitment maneuvers (data not shown), where the simulated pleural pressure behaves similarly to the recordings at the bedside, yet, the tidal volume does not.\\

\noindent\textbf{R/D and underlying time dependence} \quad
In general, the R/D processes in our model occur continuously during ventilation and behave as expected. For example, the number of open airways changes depending on the applied airway pressure. A drop in pressure results in the gradual closure of airway elements, and vice versa. On the time scale of a single breath, we observe intra-tidal R/D in our model indicated by a varying number of open airways throughout each breath (see, e.g., normal ventilation in Figure~\ref{fig:results_lf}). On the timescale of the overall ventilation protocol, we see that a reduction in driving pressure and, thus, of the tidal volume leads to less intra-tidal reopening indicated by a decreased amplitude of oscillations in the number of open airways compared to normal ventilation cycles (Figure~\ref{fig:results_deltap1}).

R/D dynamics are strongly influenced by~$\gamma$ (Figures~\ref{fig:results_lf}~-~ \ref{fig:results_rm3}), with a lower value of $\gamma$ resulting in more airways being permanently open. Moreover, changing $\gamma$ causes a change in the number of airways that transition between open/closed states with a change in ventilation modality. For example, during the second quasi-static inflation maneuver shown in Figure~\ref{fig:results_lf}, about $\sim$1.4~\% of the airways recruit in the model with the highest $\gamma$ compared to $\sim$2.2~\% of the airways in the model with the lowest~$\gamma$. We also observe this sensitivity of R/D to $\gamma$ in the rates at which airways close when the driving pressure is suddenly halved (Figures~\ref{fig:results_deltap1}), and during the period of constant PEEP immediately prior to a sustained inflation maneuver (Figure~\ref{fig:results_rm3}). The effect is most pronounced with the lowest~$\gamma$ even though the most airways remain open because of the concomitant change in $P_{\mathrm{o}}$; increasing~$\gamma$ widens the distributions of $P_{\mathrm{o}}$ and $P_{\mathrm{c}}$ and moves both to higher pressures (Figure~\ref{fig:histogram_p_open}). The effects of $\gamma$ on R/D magnitude and dynamics are thus somewhat complex.

The importance of including time dependence of R/D in the model becomes apparent when simulating the sustained-inflation maneuver. Here, the simulated gradual increases in lung volume and the number of open airways match observations well for $\gamma = 100\ \text{dyn/cm}$~(Figure~\ref{fig:results_rm3}) and show that recruitment is still continuing at the end of the maneuver. In contrast, for $\gamma = 70\ \text{dyn/cm}$, all airways reopen almost immediately after the increase in pressure, allowing the lung to reach its full capacity quickly.

Finally, our model shows permanent (de-)recruitment effects after certain maneuvers. Subsequent to the larger quasi-static inflation maneuver (Figure~\ref{fig:results_lf}), the number of open airways during normal ventilation remains slightly increased even though PEEP and driving pressure have returned to their pre-maneuver settings. This shows that temporary recruitment maneuvers can have a permanent beneficial effect on the amount of open lung. By the same token, the model also reproduces the opposite effect; when driving pressure is temporarily decreased there can be a permanent reduction in the amount of open lung (Figure~\ref{fig:results_deltap1}).\\

\begin{figure*}[htb!] 
\begin{center}
    \includegraphics[trim=0 0 0 0,clip,width=0.8\textwidth]{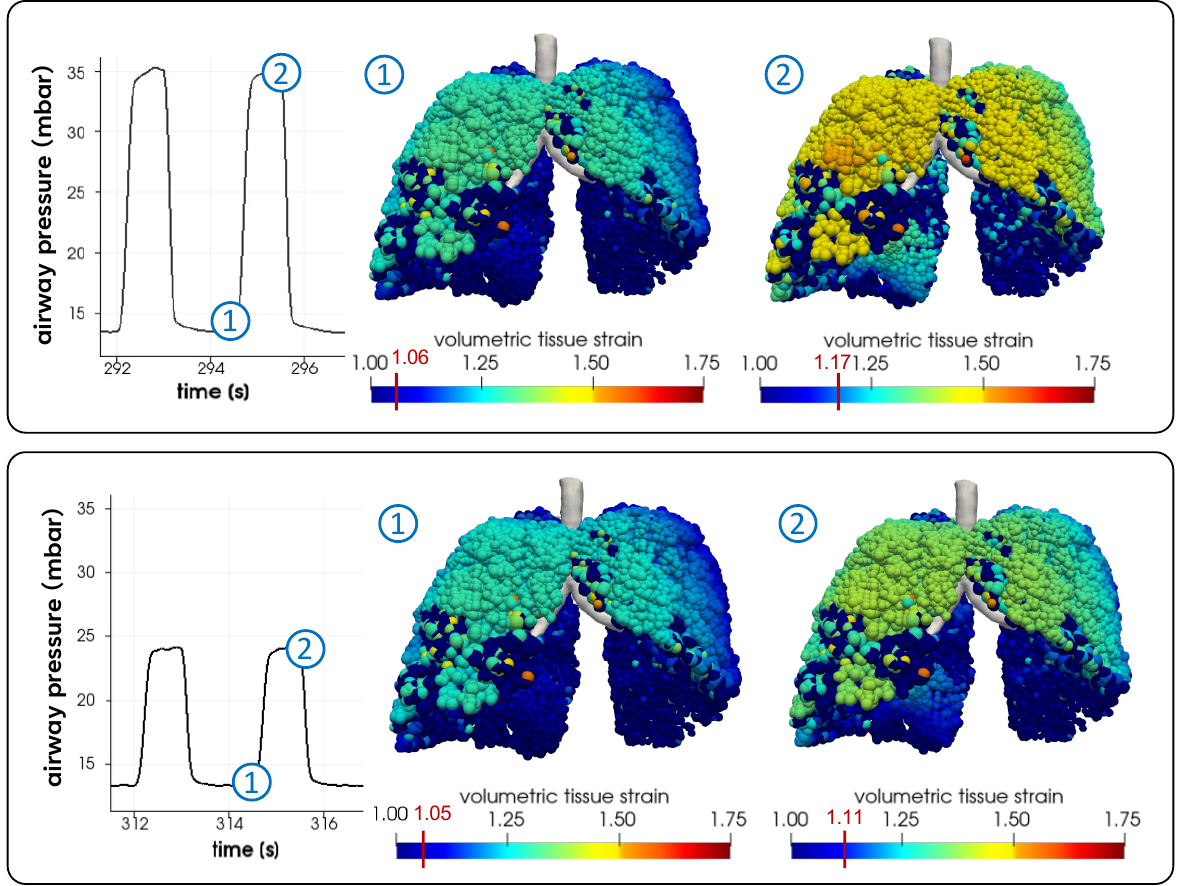}
\end{center}
\caption{Local tissue strain of normal ventilation and halved driving pressure for $\gamma = 100$~dyn/cm, at end-expiration ($t = 294.5 s$ and $t = 314.5 s$, respectively; \textcircled{1}) and at end-inspiration ($t = 295.5 s$ and $t = 315.5 s$, respectively; \textcircled{2}). The distension of an exemplary ventral terminal unit, subjected to stronger straining due to position, differs for the models with surface tension 70~dyn/cm and 130~dyn/cm from the $\gamma = 100$~dyn/cm scenario by \textcolor{black}{-4.18\% and 2.58\%} at end-expiration~(\textcircled{1}), and by \textcolor{black}{-10.41\% and 3.61\%} at end-inspiration~(\textcircled{2}) during normal ventilation, respectively; during halved driving pressure, the strain deviates by \textcolor{black}{-3.66\% and 2.25\%} at end-expiration~(\textcircled{1}), and by \textcolor{black}{-7.64\% and 2.78\%} at end-inspiration~(\textcircled{2}), respectively.}
\label{fig:results_local_strain_normal}       
\end{figure*}
\begin{figure*}[htb!] 
\begin{center}
   \includegraphics[trim=0 0 0 0 ,clip,width=0.6\textwidth]{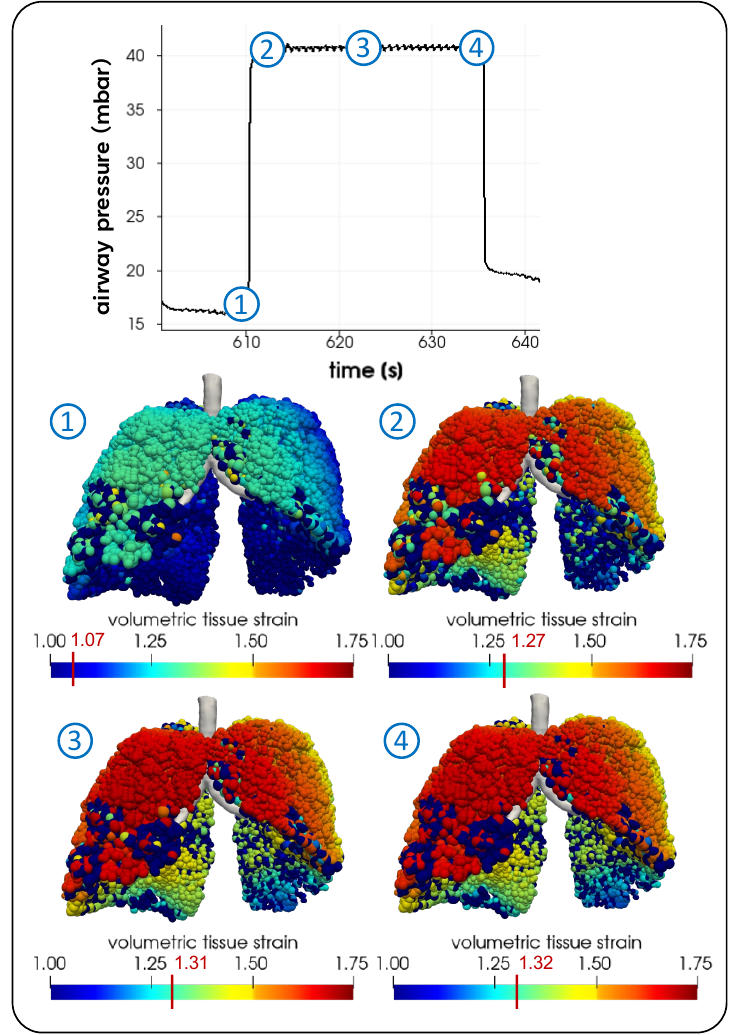}
\end{center}  
\caption{Local tissue strain during sustained-inflation maneuver for $\gamma = 100$~dyn/cm, before ($t = 610 s$, \textcircled{1}), right after onset ($t = 612 s$, \textcircled{2}), in the middle ($t = 623 s$, \textcircled{3}) and at the end ($t = 635 s$, \textcircled{4}) of the elevated pressure level. The distension of an exemplary ventral terminal unit, subjected to stronger straining due supine body, differs for the models with surface tension 70~dyn/cm and 130~dyn/cm from the $\gamma = 100$~dyn/cm scenario by \textcolor{black}{-4.91\% and 2.80\%} before~(\textcircled{1}), and \textcolor{black}{-10.82\% and 7.22\%} at the end~(\textcircled{4}) of the maneuver, respectively.}
\label{fig:results_local_strain_RM}       
\end{figure*}

\noindent\textbf{Local tissue strain} \quad
Our reduced-dimensional, yet spatially resolved, modeling approach enables us to study local tissue mechanics in the lung. Of particular interest in this regard is the strain that is experienced by different lung regions depending on local ventilation and gravitational loads. Since the model with $\gamma = 100\ \text{dyn/cm}$ shows the best agreement with clinical data (Figures~\ref{fig:results_lf}~-~\ref{fig:results_rm3}), we focus on this value of~$\gamma$ in the following.
Figures~\ref{fig:results_local_strain_normal} and~\ref{fig:results_local_strain_RM} show the tissue strains in the terminal units at selected points along the simulated airway pressure profile. At each point, an averaged global strain is determined as the ratio between current lung volume in the computational model and the end-expiratory lung volume of 3.21~l calculated from the CT image. This helps to illustrate discrepancies between the observed global strain (marked on the color bars) and the actual local strains.

Figure~\ref{fig:results_local_strain_normal} depicts the local volumetric tissue strains derived from our model during normal ventilation and at the end of the subsequent maneuver with temporarily reduced driving pressure, at two points within the breath. As expected, strains are higher during normal ventilation at peak-inspiration. However, we also see \textcolor{black}{an end-expiratory lung volume (EELV) slightly reduced by 33~ml} when driving pressure is halved. This mainly results from the viscous behavior of the terminal units; because tidal volume is reduced but the inspiration/expiration time ratio remains fixed, viscoelastic creep allow the terminal units to contract further than would be the case in normal ventilation mode. Thus, not all dynamic phenomena in the model are attributable to R/D.
Nevertheless, Figure~\ref{fig:results_local_strain_RM} also shows explicit effects of recruitment as the number of open terminal units changes throughout the course of the sustained inflation. Importantly, in contrast to the monitored global strain, the model reveals the existence of potentially harmful regional strains above 1.5~\cite{Gattinoni2017a,Protti2011}. 

Another phenomenon captured by our model is gas trapping. This occurs predominantly in regions of inhomogeneity where one can observe terminal units that have a constant strain level over a wide range of ventilation states~(Figures~\ref{fig:results_local_strain_normal} and \ref{fig:results_local_strain_RM}). It is only during a recruitment maneuver that these units finally open up.

\section{Discussion}
\label{sec:disc}

In this study, we propose an approach for comprehensively modeling the ventilatory response of a patient suffering from ARDS. Our previously described physics-based reduced-dimensional computational model based on patient-specific morphological information \cite{Roth2017a,Roth2017} is combined with an established empirical model of airway R/D dynamics~\cite{Bates2002,Massa2008,Ma2010,Smith2015,Smith2013a,Smith2015b}. We extended the R/D model by linking the critical opening pressures of closed airways to the biophysical effects of surface tension in the airway lining fluid~\cite{Naureckas1994} and the dimensions of the tracheo-bronchial tree determined from a thoracic CT image~\cite{HowatsonTawhai2000}. The CT image also allowed us to apply the R/D model specifically to diseased regions of the lung, thereby taking regional heterogeneity of lung function into account. Finally, we account for gravitational effects on the lung tissue.\\ Compared to other models in the literature~\cite{Broche2017,Knudsen2018,Ma2020,Ryans2016}, the model we have developed here is the first mechanically based and spatially resolved computational description of the entire lung that incorporates R/D dynamics and can be tailored to a specific human lung.

We parameterized the elastic behavior of the tissue elements in the model in a personalized manner by matching its behavior to that measured in a ventilated patient~\cite{Roth2017a}, focusing on a quasi-static inflation maneuver in order to calibrate the model to a wide range of pressure and volume behavior. We also accounted for the influence of the thoracic cage with a pressure boundary condition corresponding to the measured esophageal pressure that was a function of lung volume. We tested the computational model by comparing its predictions to pressure, flow and volume data measured in the mechanically ventilated ARDS patient during various respiratory maneuvers. The model recapitulated the key features of the measured data (Figs.~\ref{fig:results_lf}~-~\ref{fig:results_rm3}). The model also reproduced the dynamic R/D phenomena that led to phenomena such as gas trapping, transient opening and closing of airways, and repetitive intra-tidal recruitment, all of which may play an important role in the generation of ventilator-induced lung injury. 

A particularly novel aspect of our model is that it allows the investigation of the effects of surface tension on the dynamics of R/D. The best match of the model predictions to experimental measurement was obtained with $\gamma = 100\ \text{dyn/cm}$~(Figs.~\ref{fig:results_lf}~-~\ref{fig:results_rm3}) both in terms of lung compliance and the transient increase in lung volume seen during a sustained inflation maneuver. Since the composition of the airway liquid, and consequently the value of $\gamma$, are likely to depend on ARDS severity, choosing the value of $\gamma$ appropriately is clearly critical to model performance. This is perhaps most crucial in terms of the degree of cyclic R/D taking place within the breath, since this is likely a potent mechanism for causing lung injury. In particular, a high proportion of cyclic R/D in the model may indicate the potential for harmful shear stresses to occur at airway walls~\cite{Bilek2003} and thus might be taken as an indicator for the contribution of atelectrauma to VILI~\cite{Ghadiali2011}. Simultaneously, the model provides insight into potential risk of volutrauma by revealing local strains. 

Despite its advantages, however, the model also has some limitations. Since the composition of pathologic airway lining fluid is unknown, we assumed the same fluid properties for all diseased sites in the lung. This neglects any local changes in surface tension that might arise, for example, as a consequence of changes in the types and proportions of airway fluids that are present, or as a result of varying pathological states within the lung~\cite{Widdicombe2002j}. Nor do we consider the changes in production and secretion and the degradation of surfactant in pulmonary diseases~\cite{Seeger1993}. Even the average value of~$\gamma$ may vary among patients and between different pathological conditions and will therefore likely be difficult to determine in any specific case. Thus, the choice of the physically important parameter~$\gamma$ may have to be made on the basis of the clinical history of the patient and current conditions within the lung such as newly emerging versus long-term atelectasis along with any information that can be obtained about the ease with which collapsed lung can be recruited. Uncertainty quantification based on quantitative assessment of the effects of variations in the value of~$\gamma$ in the model will therefore likely be important. Methods of uncertainty quantification continue to advance and are proving valuable in dealing lack of knowledge in biomedical systems~\cite{Biehler2015,Wirthl2022b}.
\textcolor{black}{Another model limitation is that we only considered R/D at the airway level as a pathological phenomenon. Yet, collapse and reopening also occurs in the alveolar regions~\cite{Broche2017,Albert2009}, and the elastance or visco-elastic behavior of diseased tissue may exhibit abnormal changes~\cite{Negri2002b}. The terminal components of our model cannot mimic intermediate or diseased states of tissue distension such as might occur under pathological conditions~\cite{Broche2017,Knudsen2018} as opposed to being either fully closed or open according to Equ.~(5). Here, a novel model for alveolar R/D may be a very promising approach~\cite{Geitner2023}.}
\textcolor{black}{Last but not least, the mechanical interdependence of the individual model components resulting from tissue connectivity is limited in the lung model. On the one hand, this concerns the parenchymal tethering between airways and terminal units. We consider the influence of internal pressure in terminal units on adjacent airways and believe that the local pressure differences already have a great impact concerning the local straining behavior. However, the effect of parenchymal tethering in the classical sense, i.e., tissue connected to and pulling on airway walls is neglected. On the other hand, the interaction of the terminal units with each other could be improved. We include an indirect (global) coupling of the terminal units via the volume-dependent pressure boundary condition. Apart from that, the terminal units do not particularly influence each other at the local level. Since mechanical interdependence between terminal units, and terminal units and airways may be important for R/D dynamics~\cite{Broche2017,Mead1970,Ryans2019a}, appropriate model extensions are critical future development steps.
There exist some interesting and promising approaches in the literature~\cite{Roth2017,Ma2020,Ryans2019a}. However, further investigations are required to determine whether they adequately model the real local effects particularly with regard to the R/D phenomena and lung pathology, and how to couple them in the present model.}

Tackling these various shortcomings are promising future research objectives. In addition, using the model to make quantitative estimations of the propagation and degree of VILI caused by cyclic opening and closing and/or over-distension will be a crucial step toward optimizing mechanical ventilation in a patient-specific manner~\cite{Kollisch2018}.

\section{Conclusion}
\label{sec:concl}

In conclusion, we have introduced a novel approach to modeling the lung in a physics-based and spatially resolved manner. The model includes an empirical mechanism for R/D that is linked to airway dimensions and liquid lining properties. We applied this approach to the simulation of lung dynamics in an ARDS patient receiving mechanical ventilation in the intensive care unit. The model recapitulates the key features of the measured airway pressure, flow and volume, and provides insight into local inhomogeneity of lung function that manifests during varying pressure conditions. The model thus has the potential to elucidate spatial distributions of damaging mechanisms throughout the lung that may lead to ventilator-induced lung injury. This represents an important step toward the development of individualized therapies for the ARDS patient.


\color{black}
\appendix
\label{sec:appendix}

\section{Details of conducting airway elements}
\label{sec:app:airways}

The different types of airway resistances represent specific dissipative forces in the element. The nonlinear airway resistance~$R_{\mathrm{\mu}}$ captures the viscous and turbulent losses in the airflow and reads
\begin{equation}
    R_{\mathrm{\mu}} = \dfrac{8 \pi \mu l_{\mathrm{aw}}}{A_{\mathrm{aw}}^2}
    \left\{ 
    \begin{aligned} 
    ~\delta &\left( \dfrac{2~\mathrm{Re}}{l_{\mathrm{aw}}} \left( \dfrac{A_{\mathrm{aw}}}{\pi} \right)^{1/2} \right)^{1/2} &\mathrm{if} \quad \mathrm{Re} \geq \dfrac{l_{\mathrm{aw}}}{2\delta^2} \left( \dfrac{\pi}{A_{\mathrm{aw}}} \right)^{1/2}, \\
    &1 &\mathrm{if} \quad \mathrm{Re} < \dfrac{l_{\mathrm{aw}}}{2\delta^2} \left( \dfrac{\pi}{A_{\mathrm{aw}}} \right)^{1/2},
    \end{aligned}
    \right.
\end{equation}
where the Reynolds number Re is determined by
\begin{equation}
\mathrm{Re} = \dfrac{2 \rho \lvert Q_\mathrm{out} \rvert}{\mu \sqrt{\pi A_\mathrm{aw}}}.
\end{equation}
$A_{\mathrm{aw}}$ is the current cross-sectional area and $l_{\mathrm{aw}}$ is the length of an airway. The generation-dependent prefactor~$\delta$ was determined from experiments by~\cite{VanErtbruggen2005a} and is given in Table~\ref{tab:gen_prefactor}. $\mu = 17.9 \cdot 10^{-8}~\mathrm{mbar/s}$ is the dynamic viscosity and $\rho = 1.18~\mathrm{kg/m^{-3}}$ the density of air.
\\The inertia of the air is accounted for by the inductance~$I$ as
\begin{equation}
    I = \dfrac{\rho l_\mathrm{aw}}{A_\mathrm{aw,0}}.
\end{equation}
\\The distension of an airway is governed by the capacitance~$C$, reading
\begin{equation}
    C = \dfrac{2 \sqrt{A_\mathrm{aw}} l_\mathrm{aw}}{\eta_\mathrm{w}},
\end{equation}
where the constant geometrical and material properties of the airway wall are collected in variable
\begin{equation}
    \eta_\mathrm{w} = \dfrac{E_\mathrm{w} h_\mathrm{w} \sqrt{\pi}}{(1 - \nu^2) A_\mathrm{aw,0}}.
\end{equation}
$E_\mathrm{w}$ is the Young's modulus of the airway wall taken from~\cite{Lambert1982}, $h_\mathrm{w}$ is the airway wall thickness taken from~\cite{Montaudon2007}, $\nu$ is the Poisson's ratio set to a nearly incompressible value of 0.45, and $A_\mathrm{aw,0}$ is the unstretched cross-sectional area of an airway.
\\The distension of the airway wall is delayed by the visco-elastic resistance~$R_{\mathrm{visc}}$ inducing a phase shift $\phi_\mathrm{w} = 0.13 \mathrm{rad}$ according to 
\begin{equation}
    R_{\mathrm{visc}} = \dfrac{4 \pi \eta_\mathrm{w} T_\mathrm{w} \mathrm{tan}\phi_\mathrm{w}}{ \sqrt{A_{\mathrm{aw,0}}} l_{\mathrm{aw}}}
\end{equation}
with the time constant $T_\mathrm{w} = 2.0~\mathrm{s}$~\cite{Roth2017a}.
\\The convective resistance $R_{\mathrm{conv}}$ reads
\begin{equation}
    R_{\mathrm{conv}} =  \dfrac{2 \alpha \rho}{A_{\mathrm{aw}}^2} \left( Q_{\mathrm{out}} - Q_{\mathrm{in}} \right)
\end{equation}
with the momentum-flux correction factor~$\alpha$ according to~\cite{Sherwin2003}. For more information on the derivations of the equations for reduced-dimensional airway elements, the reader is referred to \cite{Roth2017a,Roth2017,Ismail2013,Formaggia2009,Sherwin2003,Alastruey2011}.

{
\setlength{\aboverulesep}{0pt}
\setlength{\belowrulesep}{0pt}
\setlength{\extrarowheight}{.9ex}

\begin{table}[t]
\vspace*{3cm}
    \centering
    \caption{Generation-dependent prefactor for airway resistance, taken from \cite{VanErtbruggen2005a}.}\label{tab:gen_prefactor}
    \begin{tabularx}{0.7318\linewidth}{l c c c c c c c c c}
        \toprule 
        \textbf{Generation} & 0 & 1 & 2 & 3 & 4 & 5 & 6 & 7 & >7 \\[3pt]
        \midrule
        \rowcolor{naturelight}
        $\mathbf{\delta}$ & 0.162 & 0.239 & 0.244 & 0.295 & 0.175 & 0.303 & 0.356 & 0.566 & 0.327\\[3pt]
        \bottomrule
    \end{tabularx}
\end{table}
}

\color{black}

\section*{Acknowledgements}

We gratefully acknowledge financial support by the Deutsche Forschungsgemeinschaft (DFG, German Research Foundation) in the project WA1521/26-1, by BREATHE, a Horizon 2020—ERC–2020–ADG project (grant agreement No. 101021526-BREATHE), and by NIH grant R01 HL142702.

\section*{Disclosures}

Wolfgang A. Wall is chief scientific advisor and co-founder of Ebenbuild GmbH.


\clearpage
\newpage

\vspace*{2cm}

{
\setlength{\aboverulesep}{0pt}
\setlength{\belowrulesep}{0pt}
\setlength{\extrarowheight}{.9ex}

\begin{table}[t]
\vspace*{3cm}
    \centering
    \caption{\color{Black} Patient-specific parameters of the proposed computational model.}\label{tab:fit_parameters}
    \begin{tabularx}{0.423\linewidth}{l  l  l  l}
        \toprule
        &   \textbf{Parameter}  & \textbf{Value}   & \textbf{Units} \\[3pt]
        \midrule
        \rowcolor{naturelight}
        Terminal units & $\kappa$ & \phantom{$-$}\num{3.7} & mbar \\[3pt]
        \rowcolor{white}
        (Ogden material) & $\beta$  & \num{-2.4} & - \\[3pt]
        \midrule
        \rowcolor{naturelight}
        Chest wall & $P_{\mathrm{pl,0}}$ & \phantom{$-$}\num{10.15} & mbar \\[3pt]
        \rowcolor{white}
        & $P_{\mathrm{pl,lin}}$ & \phantom{$-$}\num{9.35} & mbar \\[3pt]
        \bottomrule
    \end{tabularx}
\end{table}
}


\begin{thebibliography}{10}
\providecommand \doibase [0]{http://dx.doi.org/}%

\bibitem{Li2020}
Li X, Ma X. {Acute respiratory failure in COVID-19: Is it "typical" ARDS?}.
  {\it Critical Care} 2020\string; 24(1)\string: 1--5.
\newblock \href {\doibase 10.1186/s13054-020-02911-9} {doi:
  10.1186/s13054-020-02911-9}

\bibitem{Fan2020}
Fan E, Beitler JR, Brochard L, et al. {COVID-19-associated acute respiratory
  distress syndrome: is a different approach to management warranted?}. {\it
  The Lancet Respiratory Medicine} 2020\string; 8(8)\string: 816--821.
\newblock \href {\doibase 10.1016/S2213-2600(20)30304-0} {doi:
  10.1016/S2213-2600(20)30304-0}

\bibitem{Gattinoni2003}
Gattinoni L, Carlesso E, Cadringher P, Valenza F, Vagginelli F, Chiumello D.
  {Physical and biological triggers of ventilator-induced lung injury and its
  prevention}. {\it European Respiratory Journal, Supplement} 2003\string;
  22(47)\string: 15--25.
\newblock \href {\doibase 10.1183/09031936.03.00021303} {doi:
  10.1183/09031936.03.00021303}

\bibitem{Slutsky2013}
Slutsky AS, Ranieri VM. {Ventilator-Induced Lung Injury}. {\it New England
  Journal of Medicine} 2013\string; 369(22)\string: 2126--2136.
\newblock \href {\doibase 10.1056/NEJMra1208707} {doi: 10.1056/NEJMra1208707}

\bibitem{Beitler2016}
Beitler JR, Malhotra A, Thompson BT. {Ventilator-induced Lung Injury}. {\it
  Clinics in Chest Medicine} 2016\string; 37(4)\string: 633--646.
\newblock \href {\doibase 10.1016/j.ccm.2016.07.004} {doi:
  10.1016/j.ccm.2016.07.004}

\bibitem{Bates2018}
Bates JHT, Smith BJ. {Ventilator-induced lung injury and lung mechanics}. {\it
  Annals of Translational Medicine} 2018\string; 6(19)\string: 378--378.
\newblock \href {\doibase 10.21037/atm.2018.06.29} {doi:
  10.21037/atm.2018.06.29}

\bibitem{ARDS2000}
{The Acute Respiratory Distress Syndrome Network} . {Ventilation with Lower
  Tidal Volumes as Compared with Traditional Tidal Volumes for Acute Lung
  Injury and the Acute Respiratory Distress Syndrome}. {\it New England Journal
  of Medicine} 2000\string; 342(18)\string: 1301--1308.
\newblock \href {\doibase 10.1056/NEJM200005043421801} {doi:
  10.1056/NEJM200005043421801}

\bibitem{Kollisch2018}
Kollisch-Singule MC, Jain SV, Andrews PL, et al. Looking beyond
  macroventilatory parameters and rethinking ventilator-induced lung injury.
  {\it Journal of Applied Physiology} 2018\string; 124(5)\string: 1214-1218.
\newblock PMID: 29146685\href {\doibase 10.1152/japplphysiol.00412.2017} {doi:
  10.1152/japplphysiol.00412.2017}

\bibitem{Frerichs2017}
Frerichs I, Amato MBP, Kaam vAH, et al. Chest electrical impedance tomography
  examination, data analysis, terminology, clinical use and recommendations:
  consensus statement of the {TRanslational} {EIT} developmeNt stuDy group.
  {\it Thorax} 2017\string; 72(1)\string: 83--93.

\bibitem{Nabian2018}
Nabian M, Narusawa U. {Patient-specific optimization of mechanical ventilation
  for patients with acute respiratory distress syndrome using quasi-static
  pulmonary P-V data}. {\it Informatics in Medicine Unlocked} 2018\string;
  12(April)\string: 44--55.
\newblock \href {\doibase 10.1016/j.imu.2018.06.003} {doi:
  10.1016/j.imu.2018.06.003}

\bibitem{Morton2020}
Morton SE, Knopp JL, Tawhai MH, et al. {Prediction of lung mechanics throughout
  recruitment maneuvers in pressure-controlled ventilation}. {\it Computer
  Methods and Programs in Biomedicine} 2020\string; 197\string: 105696.
\newblock \href {\doibase 10.1016/j.cmpb.2020.105696} {doi:
  10.1016/j.cmpb.2020.105696}

\bibitem{Zhou2021}
Zhou C, Chase JG, Knopp J, et al. {Virtual patients for mechanical ventilation
  in the intensive care unit}. {\it Computer Methods and Programs in
  Biomedicine} 2021\string; 199\string: 105912.
\newblock \href {\doibase 10.1016/j.cmpb.2020.105912} {doi:
  10.1016/j.cmpb.2020.105912}

\bibitem{Sundaresan2011}
Sundaresan A, Chase JG, Shaw GM, Chiew YS, Desaive T. {Model-based optimal PEEP
  in mechanically ventilated ARDS patients in the Intensive Care Unit}. {\it
  BioMedical Engineering Online} 2011\string; 10(Mv)\string: 1--18.
\newblock \href {\doibase 10.1186/1475-925X-10-64} {doi:
  10.1186/1475-925X-10-64}

\bibitem{Sundaresan2009a}
Sundaresan A, Yuta T, Hann CE, {Geoffrey Chase} J, Shaw GM. {A minimal model of
  lung mechanics and model-based markers for optimizing ventilator treatment in
  ARDS patients}. {\it Computer Methods and Programs in Biomedicine}
  2009\string; 95(2)\string: 166--180.
\newblock \href {\doibase 10.1016/j.cmpb.2009.02.008} {doi:
  10.1016/j.cmpb.2009.02.008}

\bibitem{Bates2002}
Bates JH, Irvin CG. {Time dependence of recruitment and derecruitment in the
  lung: A theoretical model}. {\it Journal of Applied Physiology} 2002\string;
  93(2)\string: 705--713.
\newblock \href {\doibase 10.1152/japplphysiol.01274.2001} {doi:
  10.1152/japplphysiol.01274.2001}

\bibitem{Ma2011}
Ma B, Suki B, Bates JH. {Effects of recruitment/derecruitment dynamics on the
  efficacy of variable ventilation}. {\it Journal of Applied Physiology}
  2011\string; 110(5)\string: 1319--1326.
\newblock \href {\doibase 10.1152/japplphysiol.01364.2010} {doi:
  10.1152/japplphysiol.01364.2010}

\bibitem{Massa2008}
Massa CB, Allen GB, Bates JH. {Modeling the dynamics of recruitment and
  derecruitment in mice with acute lung injury}. {\it Journal of Applied
  Physiology} 2008\string; 105(6)\string: 1813--1821.
\newblock \href {\doibase 10.1152/japplphysiol.90806.2008} {doi:
  10.1152/japplphysiol.90806.2008}

\bibitem{Ma2010}
Ma B, Bates JH. {Modeling the complex dynamics of derecruitment in the lung}.
  {\it Annals of Biomedical Engineering} 2010\string; 38(11)\string:
  3466--3477.
\newblock \href {\doibase 10.1007/s10439-010-0095-2} {doi:
  10.1007/s10439-010-0095-2}

\bibitem{Smith2015}
Smith BJ, Lundblad LKA, Kollisch-Singule M, et al. {Predicting the response of
  the injured lung to the mechanical breath profile}. {\it Journal of Applied
  Physiology} 2015\string; 118(7)\string: 932--940.
\newblock \href {\doibase 10.1152/japplphysiol.00902.2014} {doi:
  10.1152/japplphysiol.00902.2014}

\bibitem{Broche2017}
Broche L, Perchiazzi G, Porra L, et al. {Dynamic Mechanical Interactions
  between Neighboring Airspaces Determine Cyclic Opening and Closure in Injured
  Lung}. {\it Critical Care Medicine} 2017\string; 45(4)\string: 687--694.
\newblock \href {\doibase 10.1097/CCM.0000000000002234} {doi:
  10.1097/CCM.0000000000002234}

\bibitem{Knudsen2018}
Knudsen L, Lopez-Rodriguez E, Berndt L, et al. {Alveolar micromechanics in
  bleomycin-induced lung injury}. {\it American Journal of Respiratory Cell and
  Molecular Biology} 2018\string; 59(6)\string: 757--769.
\newblock \href {\doibase 10.1165/rcmb.2018-0044OC} {doi:
  10.1165/rcmb.2018-0044OC}

\bibitem{Smith2013a}
Smith BJ, Grant KA, Bates JH. {Linking the development of ventilator-induced
  injury to mechanical function in the lung}. {\it Annals of Biomedical
  Engineering} 2013\string; 41(3)\string: 527--536.
\newblock \href {\doibase 10.1007/s10439-012-0693-2} {doi:
  10.1007/s10439-012-0693-2}

\bibitem{Hamlington2016}
Hamlington KL, Smith BJ, Allen GB, Bates JHT. {Predicting ventilator-induced
  lung injury using a lung injury cost function}. {\it Journal of Applied
  Physiology} 2016\string; 121(1)\string: 106--114.
\newblock \href {\doibase 10.1152/japplphysiol.00096.2016} {doi:
  10.1152/japplphysiol.00096.2016}

\bibitem{Roth2017a}
Roth CJ, Becher T, Frerichs I, Weiler N, Wall WA. {Coupling of EIT with
  computational lung modeling for predicting patient-specific ventilatory
  responses}. {\it Journal of Applied Physiology} 2017\string; 122(4)\string:
  855--867.
\newblock \href {\doibase 10.1152/japplphysiol.00236.2016} {doi:
  10.1152/japplphysiol.00236.2016}

\bibitem{Roth2017}
Roth CJ, Ismail M, Yoshihara L, Wall WA. {A comprehensive computational human
  lung model incorporating inter-acinar dependencies: Application to
  spontaneous breathing and mechanical ventilation}. {\it International Journal
  for Numerical Methods in Biomedical Engineering} 2017\string; 33(1)\string:
  1--24.
\newblock \href {\doibase 10.1002/cnm.2787} {doi: 10.1002/cnm.2787}

\bibitem{Ismail2013}
Ismail M, Comerford A, Wall WA. {Coupled and reduced dimensional modeling of
  respiratory mechanics during spontaneous breathing}. {\it International
  Journal for Numerical Methods in Biomedical Engineering} 2013\string;
  29(11)\string: 1285--1305.
\newblock \href {\doibase 10.1002/cnm.2577} {doi: 10.1002/cnm.2577}

\bibitem{Becher2021g}
Becher T, Buchholz V, Hassel D, et al. {Individualization of PEEP and tidal
  volume in ARDS patients with electrical impedance tomography: a pilot
  feasibility study}. {\it Annals of Intensive Care} 2021\string; 11(1)\string:
  89.
\newblock \href {\doibase 10.1186/s13613-021-00877-7} {doi:
  10.1186/s13613-021-00877-7}

\bibitem{HowatsonTawhai2000}
{Howatson Tawhai} M, Pullan AJ, Hunter PJ. {Generation of an Anatomically Based
  Three-Dimensional Model of the Conducting Airways}. {\it Annals of Biomedical
  Engineering} 2000\string; 28(7)\string: 793--802.
\newblock \href {\doibase 10.1114/1.1289457} {doi: 10.1114/1.1289457}

\bibitem{Weibel1963}
Weibel ER. {\it {Morphometry of the Human Lung}}.
\newblock Berlin, Heidelberg: Springer Berlin Heidelberg .
\newblock 1963

\bibitem{Horsfield1971}
Horsfield K, Dart G, Olson DE, Filley GF, Cumming G. {Models of the human
  bronchial tree}. {\it Journal of Applied Physiology} 1971\string;
  31(2)\string: 207--217.
\newblock \href {\doibase 10.1152/jappl.1971.31.2.207} {doi:
  10.1152/jappl.1971.31.2.207}

\bibitem{Majumdar2005}
Majumdar A, Alencar AM, Buldyrev SV, et al. {Relating airway diameter
  distributions to regular branching asymmetry in the lung}. {\it Physical
  Review Letters} 2005\string; 95(16)\string: 2--5.
\newblock \href {\doibase 10.1103/PhysRevLett.95.168101} {doi:
  10.1103/PhysRevLett.95.168101}

\bibitem{Formaggia2009}
Formaggia L, Quarteroni A, Veneziani A. {\it {Cardiovascular Mathematics}}.
\newblock Milano: Springer Milan .
\newblock 2009

\bibitem{Naureckas1994}
Naureckas ET, Dawson CA, Gerber BS, et al. {Airway reopening pressure in
  isolated rat lungs}. {\it Journal of Applied Physiology} 1994\string;
  76(3)\string: 1372--1377.
\newblock \href {\doibase 10.1152/jappl.1994.76.3.1372} {doi:
  10.1152/jappl.1994.76.3.1372}

\bibitem{VanOss1981}
{Van Oss} CJ, Absolom DR, Neumann AW, Zingg W. {Determination of the surface
  tension of proteins I. Surface tension of native serum proteins in aqueous
  media}. {\it BBA - Protein Structure} 1981\string; 670(1)\string: 64--73.
\newblock \href {\doibase 10.1016/0005-2795(81)90049-0} {doi:
  10.1016/0005-2795(81)90049-0}

\bibitem{Bilek2003}
Bilek AM, Dee KC, Gaver DP. {Mechanisms of surface-tension-induced epithelial
  cell damage in a model of pulmonary airway reopening}. {\it Journal of
  Applied Physiology} 2003\string; 94(2)\string: 770--783.
\newblock \href {\doibase 10.1152/japplphysiol.00764.2002} {doi:
  10.1152/japplphysiol.00764.2002}

\bibitem{Widdicombe2002j}
Widdicombe JH. {Regulation of the depth and composition of airway surface
  liquid}. {\it Journal of Anatomy} 2002\string; 201(4)\string: 313--318.
\newblock \href {\doibase 10.1046/j.1469-7580.2002.00098.x} {doi:
  10.1046/j.1469-7580.2002.00098.x}

\bibitem{Markstaller2003}
Markstaller K, Kauczor HU, Weiler N, et al. {Lung density distribution in
  dynamic CT correlates with oxygenation in ventilated pigs with lavage ARDS}.
  {\it British Journal of Anaesthesia} 2003\string; 91(5)\string: 699--708.
\newblock \href {\doibase 10.1093/bja/aeg246} {doi: 10.1093/bja/aeg246}

\bibitem{Ogden1972}
Ogden RW, A PRSL. {Large deformation isotropic elasticity – on the
  correlation of theory and experiment for incompressible rubberlike solids}.
  {\it Proceedings of the Royal Society of London. A. Mathematical and Physical
  Sciences} 1972\string; 326(1567)\string: 565--584.
\newblock \href {\doibase 10.1098/rspa.1972.0026} {doi: 10.1098/rspa.1972.0026}

\bibitem{Gattinoni2001}
Gattinoni L, Caironi P, Pelosi P, Goodman LR. {What has computed tomography
  taught us about the acute respiratory distress syndrome?}. {\it American
  Journal of Respiratory and Critical Care Medicine} 2001\string;
  164(9)\string: 1701--1711.
\newblock \href {\doibase 10.1164/ajrccm.164.9.2103121} {doi:
  10.1164/ajrccm.164.9.2103121}

\bibitem{Pelosi1994}
Pelosi P, D'Andrea L, Vitale G, Pesenti A, Gattinoni L. {Vertical gradient of
  regional lung inflation in adult respiratory distress syndrome}. {\it
  American Journal of Respiratory and Critical Care Medicine} 1994\string;
  149(1)\string: 8--13.
\newblock \href {\doibase 10.1164/ajrccm.149.1.8111603} {doi:
  10.1164/ajrccm.149.1.8111603}

\bibitem{Yoshida2018j}
Yoshida T, Amato MB, Grieco DL, et al. {Esophageal manometry and regional
  transpulmonary pressure in lung injury}. {\it American Journal of Respiratory
  and Critical Care Medicine} 2018\string; 197(8)\string: 1018--1026.
\newblock \href {\doibase 10.1164/rccm.201709-1806OC} {doi:
  10.1164/rccm.201709-1806OC}

\bibitem{West2016}
West JB, Luks A. {\it West's Respiratory Physiology - The Essentials}.
\newblock Philadelphia, United States: Wolters Kluwer.
\newblock 10th~ed. 2016.

\bibitem{Gattinoni2017a}
Gattinoni L, Marini JJ, Collino F, et al. {The future of mechanical
  ventilation: Lessons from the present and the past}. {\it Critical Care}
  2017\string; 21(1)\string: 1--11.
\newblock \href {\doibase 10.1186/s13054-017-1750-x} {doi:
  10.1186/s13054-017-1750-x}

\bibitem{Protti2011}
Protti A, Cressoni M, Santini A, et al. {Lung stress and strain during
  mechanical ventilation: Any safe threshold?}. {\it American Journal of
  Respiratory and Critical Care Medicine} 2011\string; 183(10)\string:
  1354--1362.
\newblock \href {\doibase 10.1164/rccm.201010-1757OC} {doi:
  10.1164/rccm.201010-1757OC}

\bibitem{Smith2015b}
Smith BJ, Bates JH. {Variable Ventilation as a Diagnostic Tool for the Injured
  Lung}. {\it IEEE Transactions on Biomedical Engineering} 2015\string;
  62(9)\string: 2106--2113.
\newblock \href {\doibase 10.1109/TBME.2014.2315964} {doi:
  10.1109/TBME.2014.2315964}

\bibitem{Ma2020}
Ma H, Fujioka H, Halpern D, Gaver DP. {Surfactant-Mediated Airway and Acinar
  Interactions in a Multi-Scale Model of a Healthy Lung}. {\it Frontiers in
  Physiology} 2020\string; 11(August)\string: 1--16.
\newblock \href {\doibase 10.3389/fphys.2020.00941} {doi:
  10.3389/fphys.2020.00941}

\bibitem{Ryans2016}
Ryans J, Fujioka H, Halpern D, Gaver DP. {Reduced-Dimension Modeling Approach
  for Simulating Recruitment/De-recruitment Dynamics in the Lung}. {\it Annals
  of Biomedical Engineering} 2016\string; 44(12)\string: 3619--3631.
\newblock \href {\doibase 10.1007/s10439-016-1672-9} {doi:
  10.1007/s10439-016-1672-9}

\bibitem{Ghadiali2011}
Ghadiali S, Huang Y. {Role of Airway Recruitment and Derecruitment in Lung
  Injury}. {\it Critical Reviews™ in Biomedical Engineering} 2011\string;
  39(4)\string: 297--318.
\newblock \href {\doibase 10.1615/CritRevBiomedEng.v39.i4.40} {doi:
  10.1615/CritRevBiomedEng.v39.i4.40}

\bibitem{Seeger1993}
Seeger W, G{\"{u}}nther A, Walmrath HD, Grimminger F, Lasch HG. {Alveolar
  surfactant and adult respiratory distress syndrome - Pathogenetic role and
  therapeutic prospects}. {\it The Clinical Investigator} 1993\string;
  71(3)\string: 177--190.
\newblock \href {\doibase 10.1007/BF00180100} {doi: 10.1007/BF00180100}

\bibitem{Biehler2015}
Biehler J, Gee MW, Wall WA. {Towards efficient uncertainty quantification in
  complex and large-scale biomechanical problems based on a Bayesian
  multi-fidelity scheme}. {\it Biomechanics and Modeling in Mechanobiology}
  2015\string; 14(3)\string: 489--513.
\newblock \href {\doibase 10.1007/s10237-014-0618-0} {doi:
  10.1007/s10237-014-0618-0}

\bibitem{Wirthl2022b}
Wirthl B, Brandstaeter S, Nitzler J, Schrefler BA, Wall WA. {Global sensitivity
  analysis based on Gaussian-process metamodelling for complex biomechanical
  problems}.  2022(December 2022).
\newblock \href {\doibase 10.1002/cnm.3675} {doi: 10.1002/cnm.3675}

\bibitem{Albert2009}
Albert SP, DiRocco J, Allen GB, et al. {The role of time and pressure on
  alveolar recruitment}. {\it Journal of Applied Physiology} 2009\string;
  106(3)\string: 757--765.
\newblock \href {\doibase 10.1152/japplphysiol.90735.2008} {doi:
  10.1152/japplphysiol.90735.2008}

\bibitem{Negri2002b}
Negri EM, Hoelz C, Barbas CS, Montes GS, Saldiva PH, Capelozzi VL. {Acute
  remodeling of parenchyma in pulmonary and extrapulmonary ARDS. An autopsy
  study of collagen-elastic system fibers}. {\it Pathology Research and
  Practice} 2002\string; 198(5)\string: 355--361.
\newblock \href {\doibase 10.1078/0344-0338-00266} {doi:
  10.1078/0344-0338-00266}

\bibitem{Geitner2023}
Geitner CM, Koeglmeier LJ, Becher T, et al. {Pressure- and time-dependent
  alveolar recruitment/derecruitment in a spatially resolved patient-specific
  computational model for injured human lungs}. In preparation;  2023.

\bibitem{Mead1970}
Mead J, Takishima T, Leith D. {Stress distribution in lungs: a model of
  pulmonary elasticity}. {\it Journal of Applied Physiology} 1970\string;
  28(5)\string: 596--608.
\newblock \href {\doibase 10.1152/jappl.1970.28.5.596} {doi:
  10.1152/jappl.1970.28.5.596}

\bibitem{Ryans2019a}
Ryans JM, Fujioka H, Gaver DP. {Microscale to mesoscale analysis of parenchymal
  tethering: The effect of heterogeneous alveolar pressures on the pulmonary
  mechanics of compliant airways}. {\it Journal of Applied Physiology}
  2019\string; 126(5)\string: 1204--1213.
\newblock \href {\doibase 10.1152/japplphysiol.00178.2018} {doi:
  10.1152/japplphysiol.00178.2018}

\bibitem{VanErtbruggen2005a}
{Van Ertbruggen} C, Hirsch C, Paiva M. {Anatomically based three-dimensional
  model of airways to simulate flow and particle transport using computational
  fluid dynamics}. {\it Journal of Applied Physiology} 2005\string;
  98(3)\string: 970--980.
\newblock \href {\doibase 10.1152/japplphysiol.00795.2004} {doi:
  10.1152/japplphysiol.00795.2004}

\bibitem{Lambert1982}
Lambert RK, Wilson TA, Hyatt RE, Rodarte JR. {A computational model for
  expiratory flow}. {\it Journal of Applied Physiology Respiratory
  Environmental and Exercise Physiology} 1982\string; 52(1)\string: 44--56.
\newblock \href {\doibase 10.1152/jappl.1982.52.1.44} {doi:
  10.1152/jappl.1982.52.1.44}

\bibitem{Montaudon2007}
Montaudon M, Desbarats P, Berger P, Dietrich dG, Marthan R, Laurent F.
  {Assessment of bronchial wall thickness and lumen diameter in human adults
  using multi-detector computed tomography: Comparison with theoretical
  models}. {\it Journal of Anatomy} 2007\string; 211(5)\string: 579--588.
\newblock \href {\doibase 10.1111/j.1469-7580.2007.00811.x} {doi:
  10.1111/j.1469-7580.2007.00811.x}

\bibitem{Sherwin2003}
Sherwin SJ, Franke V, Peir{\'{o}} J, Parker K. {One-dimensional modelling of a
  vascular network in space-time variables}. {\it Journal of Engineering
  Mathematics} 2003\string; 47(3-4)\string: 217--250.
\newblock \href {\doibase 10.1023/B:ENGI.0000007979.32871.e2} {doi:
  10.1023/B:ENGI.0000007979.32871.e2}

\bibitem{Alastruey2011}
Alastruey J, Khir AW, Matthys KS, et al. {Pulse wave propagation in a model
  human arterial network: Assessment of 1-D visco-elastic simulations against
  in vitro measurements}. {\it Journal of Biomechanics} 2011\string;
  44(12)\string: 2250--2258.
\newblock \href {\doibase 10.1016/j.jbiomech.2011.05.041} {doi:
  10.1016/j.jbiomech.2011.05.041}

\end{thebibliography}
\end{document}